\documentclass[preprint,12pt,3p,superscriptaddress]{elsarticle}
\usepackage{amssymb}
\usepackage{lineno}
\usepackage[utf8]{inputenc}
\usepackage[T1]{fontenc}
\usepackage{epsfig}
\usepackage{amsmath}
\usepackage{amsfonts}
\usepackage{amssymb}
\usepackage{color}
\usepackage{graphicx}
\usepackage{url,hyperref}
\usepackage{latexsym}
\usepackage{longtable}
\usepackage{float}
\usepackage{epstopdf}
\usepackage{fullpage}
\usepackage{times}
\usepackage{fancyheadings}
\journal{Annals of Physics}
\begin{document}
\begin{frontmatter}
\title{Collective motion in prolate $\gamma$-rigid nuclei within minimal length concept via a quantum perturbation method}
\author{M. Chabab}
\author{A. El Batoul}
\author{A. Lahbas}
\author{M. Oulne\corref{cor1}}
\ead{oulne@uca.ma}
\address{High Energy Physics and Astrophysics Laboratory, Department of Physics, Faculty of Sciences Semlalia, Cadi Ayyad University P.O.B 2390, Marrakesh 40000, Morocco.}
\cortext[cor1]{corresponding author}
\date{\today}
\begin{abstract}
\noindent
Based on the minimal length concept, inspired by Heisenberg algebra, a closed analytical formula is derived for the energy spectrum of the prolate $\gamma$-rigid Bohr-Mottelson Hamiltonian of nuclei, within a quantum perturbation method (QPM), by considering a scaled Davidson potential in $\beta$ shape variable. In the resulting solution, called X(3)-D-ML, the ground state and the first $\beta$-band are all studied as a function of the free parameters. The fact of introducing the minimal length concept with a QPM makes the model very flexible and a powerful approach to describe nuclear collective excitations of a variety of vibrational-like nuclei. The introduction of scaling parameters in the Davidson potential enables us to get a physical minimum of this latter in comparison with previous works. The analysis of the corrected wave function, as well as the probability density distribution, shows that the minimal length parameter has a physical upper bound limit.
\end{abstract}
\begin{keyword}
Bohr-Mottelson model, $\gamma$-rigid axial symmetry, critical point symmetries, minimal length, Davidson potential, quantum perturbation method.
\PACS 21.60.Ev \sep 21.60.Fw \sep 21.10.Re
\end{keyword}
\end{frontmatter}
\section{Introduction}
The introduction of critical point symmetries concept \cite{b1,b2,b3,b4,b5,b6,b7,b8}, describing nuclei at points of shape-phase transitions between different limiting symmetries, was originally suggested by Iachello\cite{b1,b2,b3}. It is still one of the hot topics in nuclear structure physics. Moreover, a much development, in this direction, has been mainly accomplished by both phenomenological models: the interacting boson model (IBM)\cite{b9} as well as  the  hydrodynamical Bohr-Mottelson model\cite{b10,b11}. The latter is very useful in describing the rotation and vibrations for quadrupole deformed nuclei. Particularly, for the shape evolution and phase transitions\cite{b12}, it is a powerful tool for inspecting the critical point symmetries like E(5)\cite{b1}, X(5)\cite{b2}, Y(5)\cite{b3} and Z(5)\cite{b13}, which describe the nuclei situated in the critical points of the shape phase transitions from spherical vibrator [U(5)] to a $\gamma$-unstable [O(6)] nuclei, from spherical vibrator to prolate rotor [SU(3)], from axial rotor to triaxial rotor, and from prolate rotor to oblate rotor, respectively. Going further, most of the critical-point symmetries mentioned above are located at each vertex in a diagram of what is terminologically called the Casten triangle and are considered as verifiable benchmarks for the experiment thanks to their parameter-free solutions. As a matter of fact, the details of these physical situations can be found in Refs. \cite{b4,b14}. Other efforts have also been directed to special realizations in the framework of the Bohr-Mottelson model and its extensions\cite{b15,b16,b17,b18,b19,b20,b21,b22,b23,b24,b25} where the collective shape variables or inertial parameters are imposed by some constraints. From a structural point of view, the collective Bohr Hamiltonian induces, however, a coupling of the $\beta$, $\gamma$, and rotational degrees of freedom, thereby yielding a rich set of physical phenomena. For example, by fixing the $\gamma$ variable to be equal to zero ($\gamma=0$), one obtains the $\gamma$-rigid version of the critical point symmetry X(5) being called X(3). Not long ago, new improved versions of the standard X(5) and X(3) symmetries being called X(5)-ML and X(3)-ML have been elaborated by introducing for the first time the minimal length concept in nuclear structure\cite{b15}. We recall that in this work, we started by modifying the momentum operator according to the requirements of the finite length theory, to obtain the relevant collective Bohr Hamiltonian through the Pauli-Podolsky prescription, and which we solved by using standard techniques, reaching analytic expressions for the spectra and the corresponding wave functions. 
	On the other hand, to highlight the philosophy of the minimal length, we first describe its principles. Basically, the introduction of this elementary length is equivalent to an additional uncertainty in the measurement of the position, so that the minimum uncertainty can never be zero. Additionally, the minimal length is particularly useful to solve problems characterized by anomalies owing singularities at small distances. Thus, several studies in string theory\cite{b26,b27,b28} and quantum gravity\cite{b29,b30,b31} in the view of Heisenberg algebra propose a small corrections to the Heisenberg uncertainty relation of the form 
	\begin{equation}
	(\Delta X)(\Delta P) \succeq (\hbar/2)\left[1+\alpha(\Delta P)^2+\cdots\right],
	\label{E1}
	\end{equation}
	Therefore this correction results in the modification of the canonical commutation relation between the position operator  and momentum operator which becomes:
	\begin{equation}
	\left[X,P\right]=i\hbar\left(1+\alpha P^2+\cdots\right)
	\label{E2}
	\end{equation}
	According to this assumption, the space parameters of the model will be disturbed due to a strong coupling between the minimal length parameter and the degrees of freedom associated with the model. 
	Therefore, our aim in the present work is to use the idea of minimal length, as in Ref. \cite{b15},but this time in the framework of X(3) model and a Davidson potential for the collective shape variable, i.e., $\beta$. The model is conventionally called X(3)-D-ML in connection with the standard X(3)-ML model\cite{b15}. In addition to the points mentioned above, the following comments apply:
	\begin{itemize}
		\item In practical interest, the Hamiltonian of the system is not soluble analytically for the Davidson-type potential. However, the quantum perturbation theory one of its familiar forms, dubbed the quantum perturbation method (QPM), is used to obtain approximate solutions for all values of angular momentum L.
		\item In the standard Davidson potential, scaling parameters are introduced in order to obtain  physical values for its minimum.
		\item To test the applicability of the Davidson potential  in the description of nuclear spectra, we have fitted some nuclei having  the mass number $100\preceq A\preceq 220$ and   the  observed signature $1.97\preceq R_{4/2} \preceq3.2$. Furthermore, the experimental realization of the model is found in the following nuclei: ${}^{100}$Mo, ${}^{100-102}$Pd, ${}^{116}$Te, ${}^{130}$Xe, ${}^{148}$Sm, ${}^{150}$Nd, ${}^{152}$Gd, ${}^{154}$Er, ${}^{176-180}$Os, ${}^{184}$Os, ${}^{180-186}$Pt and ${}^{220}$Th.
	\end{itemize}
	In completion of the  current study, a phenomenological interpretation of the model is proposed in order to find a  physical meaning for description of candidate nuclei.
	\section{Theoretical underpinnings of the  Model}
	It's well known that in the collective structure of atomic nuclei, particulary for a spherical system, the quadrupole deformation is the fundamental mode of deformation which can be described by a set of five amplitudes that form the components of a spherical tensor. Typically, a set of five amplitudes that form the components of a spherical tensor. In the framework of Bohr-Mottelson model, these tensorial components are the two dynamical variables $\beta$ and $\gamma$ plus the three Euler angles  $\theta_i(i=1,2,3)$. In the same context, the classical expression for the rigid-body kinetic energy associated with the rotation and surface deformations of	a nucleus has the form \cite{b2,b32}
	\begin{equation}
	\hat{T} = \frac{1}{2}\sum_{k=1}^{3} {\cal J}_k\, \omega^{\prime2}_k +
	\frac{B_m}{2}\,(\dot{\beta}^2+\beta^2 \dot{\gamma}^2), 
	\label{E3}
	\end{equation}
	where $B_m$ is the 
	mass parameter, 
	\begin{equation}
	{\cal J}_k = 4B_m\beta^2 \sin^2\bigl(\gamma - {\textstyle\frac{2}{3}}\pi k\bigr)
	\label{E4}
	\end{equation}
	are the three principal irrotational moments of inertia, 
	and $\omega^\prime_k$ ($k=1$, 2, 3) are the components of the angular  frequencies
	on the body-fixed $k$-axes, which can be expressed in terms of the time derivatives of the Euler angles, 
	\begin{eqnarray}
	\omega^\prime_1 &=& -\sin\theta \cos\psi\,\dot{\phi} + \sin\psi\,\dot{\theta}, 
	\nonumber\\
	\omega^\prime_2 &=& \sin\theta \sin\psi\,\dot{\phi} + \cos\psi\,\dot{\theta},\\
	\omega^\prime_3 &=& \cos\theta\,\dot{\phi} + \dot{\psi}. \nonumber
	\label{E5}
	\end{eqnarray}
	It should be noted however that by imposing some constraints on the kinetic energy, one can reduce the number of degrees of freedom and therefore obtain a different Hamiltonian forms. In this work, we consider the case when the $\gamma$  degree of freedom is frozen to $0^\circ$. Therefore, we aimed to reveal the minimal length effect on energy spectrum in the context of $\gamma$ rigid nuclei. So, by employing the mathematical formulation, including the minimal length concept, presented in the original paper\cite{b15}, the collective  equation of eigenstates, up to the first order of $\alpha$, is written as follows:
	\begin{equation}
	\Biggl[-\frac{\hbar^2}{2B_m}\Delta +\frac{\alpha\hbar^4}{B_m}\Delta^2+ U(\beta)-E_{n,L} \Biggr]\Psi(\beta,\theta,\phi)=0,
	\label{E6}
	\end{equation}
	where 
	\begin{align}
	\Delta = \Biggl[\frac{1}{\beta^2}
	\frac{\partial}{\partial\beta}\beta^2\frac{\partial}{\partial\beta} +
	\frac{\Delta _\Omega}{3\beta^2} \Biggr] 
	\label{E7}
	\end{align}
	and $\Delta_\Omega$ is the angular part of the Laplace operator
	\begin{equation}
	\Delta_\Omega =\left[ \frac{1}{\sin\theta}\frac{\partial}{\partial\theta}\sin\theta
	\frac{\partial}{\partial\theta} + \frac{1}{\sin^2\theta}
	\frac{\partial^2}{\partial\phi^2}\right]. 
	\label{E8}
	\end{equation}
	This equation can be simplifed by introducing an auxiliary wave function\cite{b15}:
	\begin{equation}
	\Psi(\beta,\theta,\phi)=\left[1+2\alpha\hbar^2\Delta\right]\Phi(\beta,\theta,\phi)
	\label{E9}
	\end{equation}
	leading to the following differential equation,
	\begin{eqnarray}
	\Biggl[\left(1+4B_m\alpha\left(E_{n,L}-U(\beta)\right)\right)\Delta+\frac{2B_m}{\hbar^2}\bigg(E_{n,L}-U(\beta)\bigg)\Biggr]\Phi\left(\beta,\theta,\phi\right)=0 
	\label{E10}
	\end{eqnarray}
	In addition, separation of variables can be achieved by assuming the wave function to be of the form	
	\begin{equation}
	\Phi(\beta,\theta,\phi) = F_{n_{\beta}}(\beta)\, Y_{LM}(\theta,\phi),
	\label{E11}
	\end{equation}
	where $Y_{LM}(\theta,\phi)$ are the spherical harmonics. Then the angular part leads to the equation
	\begin{equation}
	\Delta_\Omega Y_{LM}(\theta,\phi) =- L(L+1)Y_{LM}(\theta,\phi),
	\label{E12}
	\end{equation}
	Here, $L$ is the angular momentum quantum number, while  the radial part $F(\beta)$ obeys to:
	\begin{equation}
	\Biggl[\frac{1}{\beta^2}
	\frac{d}{d\beta}\beta^2\frac{d}{d\beta}
	-\frac{L(L+1)}{3\beta^2} + \frac{2B}{\hbar^2}\bar{K}(E,\beta)\Biggr]F_{n_{\beta}}(\beta) = 0.
	\label{E13}
	\end{equation}
	with
	\begin{equation}
	\bar{K}(E_{n,L},\beta)=\left(\frac{E_{n,L}-U(\beta)}{\left(1+4B_m\alpha\left(E_{n,L}-U(\beta)\right)\right)}\right)
	\label{E14}
	\end{equation}
	and $n_{\beta}$ is the radial quantum number.
	Thanks to the smallness of the parameter $\alpha$, by expanding Eq. (\ref{E14}) in power series of $\alpha$, one can obtain different order approximations of the standard model X(3)-ML. At the first order approximation, as it has been done recently in \cite{b16}, Eq. (\ref{E14}) becomes:
	\begin{eqnarray}
	\bar{K}(E_{n,L},\beta)&\approx\left(E_{n,L}-U(\beta)\right)
	\left(1-4B_m\alpha\left(E_{n,L}-U(\beta)\right)\right)\nonumber\\
	&=E_{n,L}-U(\beta)-4B_m\alpha\left(E_{n,L}-U(\beta)\right)^2\nonumber\\
	\label{E15b}
	\end{eqnarray}
	This approximation only provides an approximate version and not an alternative one to X(3)-ML.
	In what concerns the $\beta$ degree of freedom, we will consider the Davidson like potential. The latter is chosen to be of the following form:
	\begin{equation}
	U(\beta)=a\beta^2+\frac{b}{\beta^2}, \ \beta_0=\left(\frac{b}{a}\right)^{1/4}
	\label{E15}
	\end{equation}
	where $a$ and $b$ are two free scaling parameters, and $\beta_0$ represents the position of the minimum of the potential. The special case of $b=0$ ($\beta_0=0$) corresponds to the  simple harmonic oscillator. 
	The $\beta$ differential equation (\ref{E13}) was solved exactly, with an infinite square well like potential, within the standard method, but for this potential it's not exactly solvable. Thus an approximate method is required. For this purpose, we adopt the quantum perturbation method (QPM)\cite{b34} which is widely used in quantum perturbation theory.
	\subsection{Treatment of the ordinary cas  $\alpha=0$ within AIM}
	In what follows, it is preferable to write equation (\ref{E13}) in a Schr\"{o}dinger picture. This is realized
	by changing the wave function as $F_{n_{\beta}}(\beta)=\chi_{n_{\beta}}(\beta)/\beta^2$. However one obtains an equation which resembles the radial Schr\"{o}dinger equation for an isotropic Harmonic Oscillator acting in three-dimensional space:
	\begin{equation}
	\frac{d^2\chi_{\beta}(\beta)}{d\beta^2}+	\Big[\epsilon-\omega\beta^2-\frac{m_{L,b}\left(m_{L,b}+1\right)}{\beta^2}\Big]\chi_{n_{\beta}}(\beta) = 0.
	\label{E16}
	\end{equation}
	with,
	\begin{align}
	&\epsilon=\frac{2B_mE_{n,L}}{\hbar^2}, \ \omega=\frac{2B_ma}{\hbar^2}
	\label{E17_1}
	\end{align}
	and
	\begin{align}
	m_{L,b}(m_{L,b}+1)=\frac{2B_mb}{\hbar^2}+\frac{L(L+1)}{3}
	\label{E17}
	\end{align}
	To solve this differential equation via the asymptotic iteration method (AIM)\cite{b35,b36}, we propose the
	following ansatz:
	\begin{equation}
	\chi_{n_{\beta}}(\beta)=\beta^{\left(1+m_{L,b}\right)}e^{-\frac{\sqrt{\omega}}{2}\beta^2}\xi_{n_{\beta}}(\beta)
	\label{E18}
	\end{equation} 
	Thus we obtain,
	\begin{equation}
	\frac{d^2\xi_{\beta}(\beta)}{d\beta^2}+\Big[\frac{2p}{\beta}-4q\beta\Big]\frac{d\xi_{n_{\beta}}(\beta)}{d\beta}+\Big[\epsilon-2q\left(1+2p\right)\Big]\xi_{n_{\beta}}(\beta) = 0.
	\label{E19}
	\end{equation}
	where we have used the parametrization:
	\begin{equation}
	p=1+m_{L,b},\ q=\sqrt{\omega}/2.
	\label{E20}
	\end{equation}
	After calculating $\lambda_0$ and $s_0$, by means of the recurrence relations of equation (2.3) given in Ref. \cite{b36}, we get the generalized formula of the reduced energy from
	the roots of the quantization condition(Eq. (2.6) in Ref. \cite{b36}) as follows:
	\begin{equation}
	\epsilon=q\left[2+4p+8n_{\beta}\right], n_{\beta}=0,1,2,\cdots,
	\label{E20_1}
	\end{equation}
	from which, we obtain the energy spectrum :
	\begin{equation}
	E_{n_{\beta},L}^{(0)}=\frac{\hbar^2}{2B_m}\epsilon=\sqrt{\frac{\hbar^2}{2B_m}a}\Big[3+4n_{\beta}+2m_{L,b}\Big]
	\label{E21}
	\end{equation}
	From equation (\ref{E17}), we get $m_{L,b}$ as a function of the total angular momentum $L$ and the parameter $b$ :
	\begin{equation}
	m_{L,b}=-\frac{1}{2}+\frac{1}{2}\sqrt{1+4\left(\frac{2B_mb}{\hbar^2}+\frac{L(L+1)}{3}\right)}
	\label{E22}
	\end{equation}
	The physical solutions to the differential equation (\ref{E13}) are obtained as:
	\begin{eqnarray}
	F_{n_{\beta}}(\beta)=& N_{n_{\beta},L}\cdot\beta^{(p-2)}e^{-q\beta^2}{}_1\mathcal{F}_{1}\left(-n_{\beta};p+\frac{1}{2};2q\beta^2\right)\nonumber\\=& N_{n_{\beta},L}\cdot\beta^{(p-2)}e^{-q\beta^2}\mathcal{L}_{n_{\beta}}^{(p-\frac{1}{2})}\left(2q\beta^2\right)
	\label{E23}
	\end{eqnarray}
	where $\mathcal{L}_{n}^{(x)}(t)$ denotes the associated Laguerre polynomials and $N_{n_{\beta},L}$ is a normalization constant to be determined later.
	\subsection{Treatment of the case $\alpha\ne0$ within QPM}
	Here we treat the additional term $(\alpha\hbar^4/B_m)\Delta^2$ shown in equation (\ref{E6}) as a perturbation and then estimate its effect on the energy spectrum  up to the first order of the perturbation theory. Hence, the energy spectrum can be written as:
	\begin{equation}
	E_{n_{\beta},L}=E_{n_{\beta},L}^{(0)}+\Delta E_{n_{\beta},L},
	\label{E24}
	\end{equation}
	where $E_{n_{\beta},L}^{(0)}$ are the unperturbed levels corresponding to the
	eigenfunctions $\psi_{n_{\beta},L}^{(0)}(\beta)$, solutions to the ordinary
	Schr\"{o}dinger equation, and $\Delta E_{n_{\beta},L}$ is the correction induced by
	the minimal length, given by:
	\begin{equation}
	\Delta E_{n_{\beta},L}=\alpha\frac{\hbar^4}{B_m}\langle\psi_{n_{\beta},L}^{(0)}\left\vert
	\Delta^{2}\right\vert \psi_{n_{\beta}^{\prime},L^{\prime}}^{(0)}\rangle
	\label{E25}
	\end{equation}
	which can be expressed as,
	\begin{eqnarray}
	\Delta E_{n_{\beta},L}=4B_m\alpha\bigg[\left(  E_{n_{\beta},L}^{(0)}\right)^{2}
	-2E_{n_{\beta},L}^{(0)}\langle\psi_{n_{\beta},L}^{(0)}\left\vert U(\beta)\right\vert \psi_{n_{\beta},L}^{(0)}\rangle
	+\langle\psi_{n_{\beta},L}^{(0)}\left\vert U(\beta)^2\right\vert \psi_{n_{\beta},L}^{(0)}\rangle\bigg].
	\label{E26}
	\end{eqnarray}
	After substituting  the Davidson potential (\ref{E15})  into Eq. (\ref{E26}), one obtains
	\begin{eqnarray}
	\Delta E_{n_{\beta},L}=& 4B_m\alpha\bigg[\left(  E_{n_{\beta},L}^{(0)}\right)^{2}+2ab
	-2E_{n_{\beta},L}^{(0)}\left(a\overline{\beta^2}+b\overline{\beta^{-2}}\right)+ \left(a^2\overline{\beta^4}+b^2\overline{\beta^{-4}}\right)\bigg].
	\label{E27}
	\end{eqnarray}
	where $\overline{\beta^{t}} (t=2,-2,4,-4)$ are expressed as follows:
	\begin{align}
	&\overline{\beta^{2}}=\frac{4n_{\beta}+2m_{L,b}+3}{4q},\nonumber\\
	&\overline{\beta^{-2}}=\frac{4q}{2m_{L,b}+1},\nonumber\\
	&\overline{\beta^{4}}=\frac{4m_{L,b}^2+24n_{\beta}m_{L,b}+24n_{\beta}^2+16m_{L,b}+36n_{\beta}+15}{16q^2},\nonumber\\
	&\overline{\beta^{-4}}=\frac{16q^2\left(4n_{\beta}+2m_{L,b}+3\right)}{\left(2m_{L,b}+3\right)\left(4m_{L,b}^2-1\right)}.
	\label{E28}
	\end{align}
	Details of $\overline{\beta^{t}}$ calculations are given in Appendix A, while $m_{L,b}$ is given by Eq. (\ref{E22}).
	The obtained formula given by Eq. (\ref{E27}) is the main result of this work. It allows us to investigate the effect of the minimal length  as well as the scaling parameters of Davidson potential  on the  energy levels of a given nucleus. Besides, we can remark that the minimal length correction carries new terms in  the energy spectrum with respect to the undeformed case. It is obvious that in quantum mechanical problems, the wave function is as important as the energy levels. Therefore, the next step is to calculate the corrected wave function of our model using the same method. By employing QPM, the first-order correction to the wave function is given by :
	\begin{equation}
		F^{Corr}_{n_{\beta}}(\beta)=F_{n_{\beta}}(\beta)+\sum_{k\neq n_{\beta}}\left[\frac{\int_{0}^{\infty}\beta^2F_{k}(\beta)\vartheta(n_{\beta},\alpha,a,b,E_{n_{\beta},L}^{(0)})F_{n_{\beta}}(\beta)d\beta}{E_{n_{\beta},L}^{(0)}-E_{k,L}^{(0)}}\right]F_{k}(\beta)
		\label{E28_a}
	\end{equation}
	with
	\begin{eqnarray}
	\vartheta(n,\alpha,a,b,E_{n,L}^{(0)})=& 4B_m\alpha\bigg[\left(  E_{n,L}^{(0)}\right)^{2}+2ab
	-2E_{n,L}^{(0)}\left(a\beta^2+b\beta^{-2}\right)+ \left(a^2\beta^4+b^2\beta^{-4}\right)\bigg].
	\label{E28_b}
	\end{eqnarray}
    Having the corrected wave function, we can also calculate the probability density distribution,
	\begin{equation}
	\rho_{n_{\beta},L}(\beta)=\beta^2|F^{Corr}_{n_{\beta}}(\beta)|^2
	\label{E28_c}
	\end{equation}		
	\section{Numerical examination and Discussion}
\begin{figure}[!h]
		\centering
		\rotatebox{0}{\includegraphics[height=50mm]{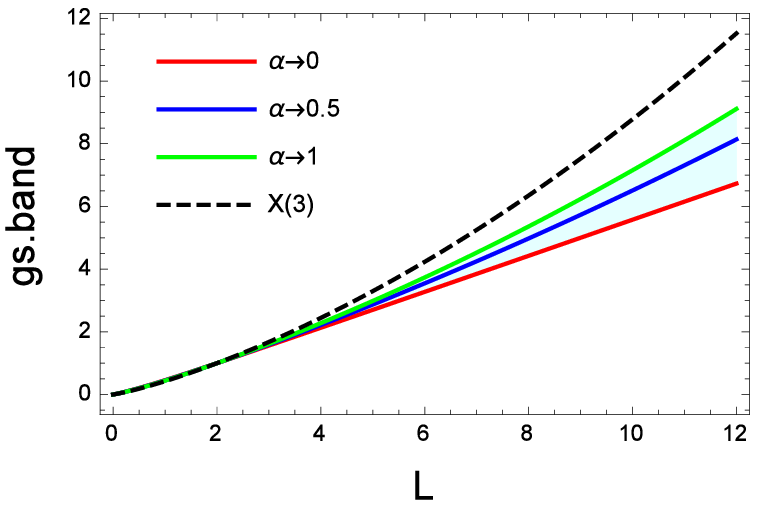}} 
		\rotatebox{0}{\includegraphics[height=50mm]{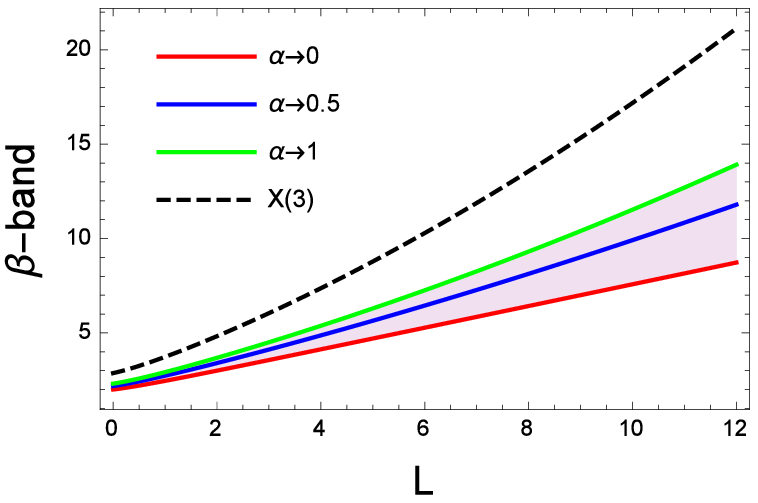}}
		\caption{ The region allowed by the X(3)-D-ML model of the energy spectrum with respect to the ground state and the $\beta$ bands, normalized to the energy of the first excited state  are plotted as a function of angular momentum L using the Davidson potential with 
the following parameters: $a=0.004$ and $b=0$. The X(3) prediction is also shown for comparison. The region allowed by our model is located below X(3) model for ground state band as well as $\beta$ band .}
		\label{Fig_1}
	\end{figure}

\begin{figure}[!h]
		\centering
		\rotatebox{0}{\includegraphics[height=50mm]{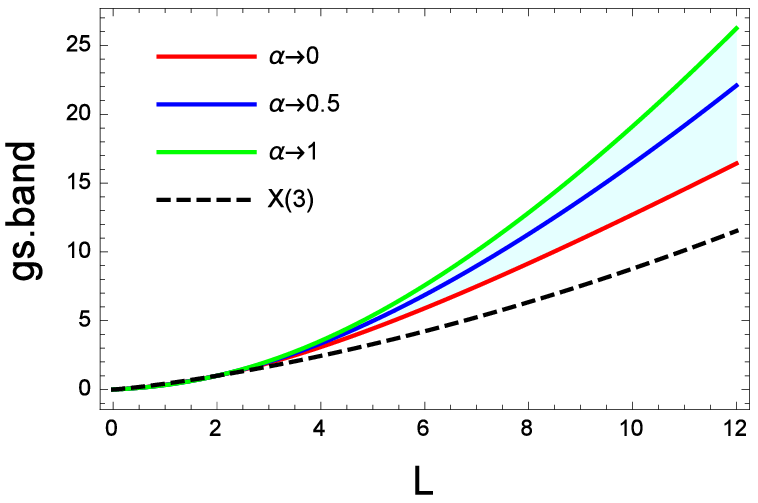}} 
		\rotatebox{0}{\includegraphics[height=50mm]{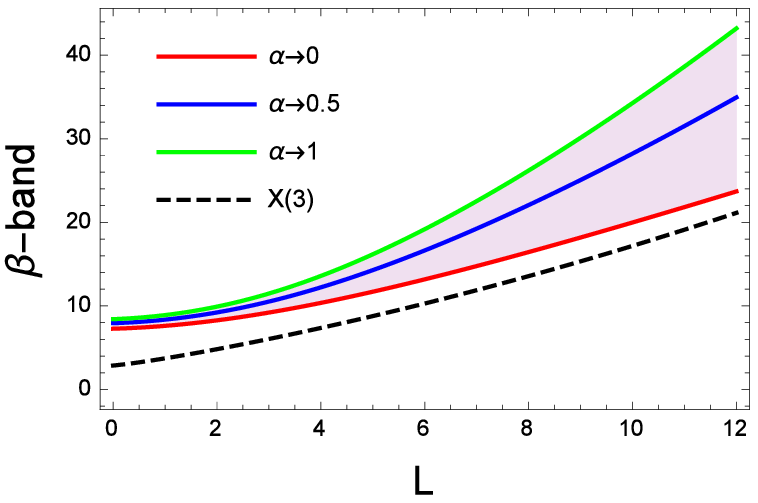}}
		\caption{ The same as in Fig. \ref{Fig_1}, but for $a=0.004$ and $b=6$. In this case, the allowed region is located above X(3) for ground state band and $\beta$ band.}
		\label{Fig_2}
	\end{figure}

\begin{figure}[!h]
	 	\centering	
	 	\rotatebox{0}{\includegraphics[height=50mm]{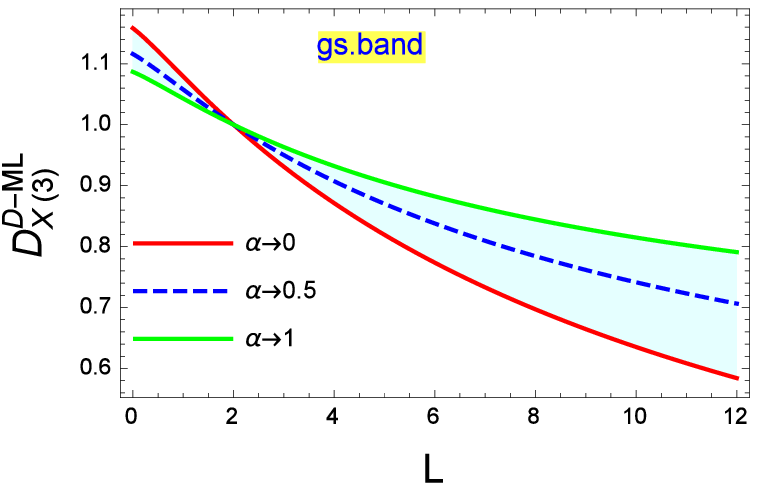}} 
	 	\rotatebox{0}{\includegraphics[height=50mm]{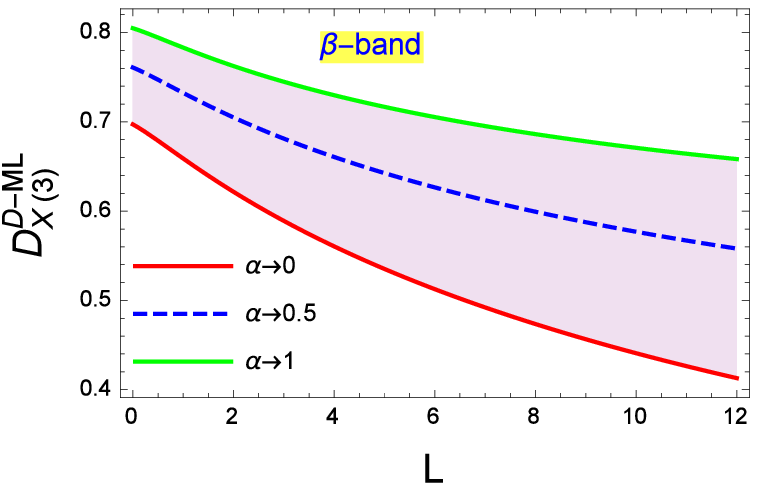}}
	 	\caption{The deviation D$_{X(3)}^{D-ML}$ which encloses the existence region of the model for the energy spectrum of ground state and $\beta$-band  compared to the energies of X(3) model are visualized as a function of angular momentum L using the Davidson potential with the following parameters: $a=0.004$ and $b=0$.}
	 	\label{Fig_3}
	 \end{figure}

\begin{figure}[!h]
	 	\rotatebox{0}{\includegraphics[height=50mm]{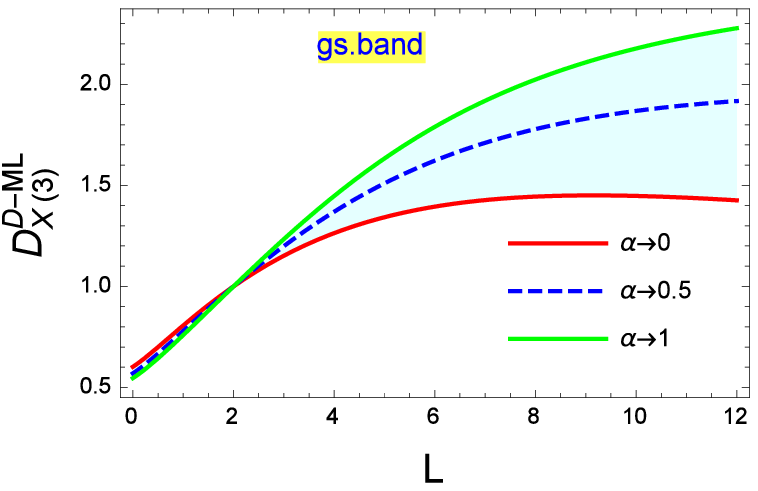}} 
	 	\rotatebox{0}{\includegraphics[height=50mm]{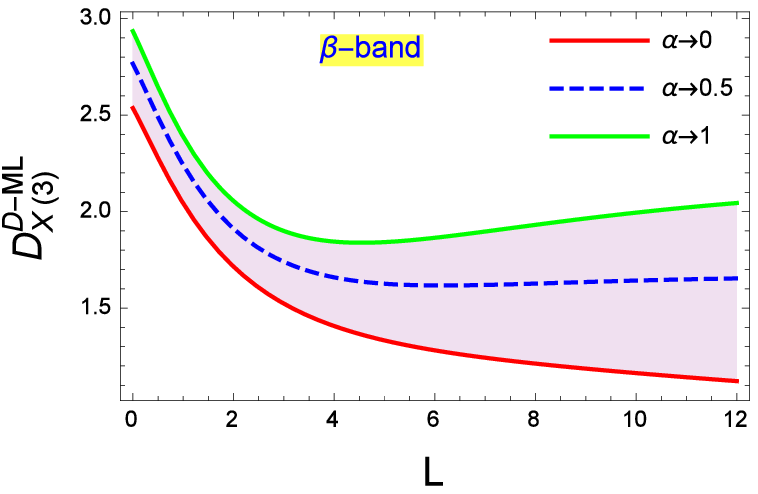}}
	 	\caption{The same as in Fig. \ref{Fig_3}, but for $a=0.004$ and $b=6$.}
	 	\label{Fig_4}
	 \end{figure}

\begin{figure}[!h]
		\centering	
		\rotatebox{0}{\includegraphics[height=50mm]{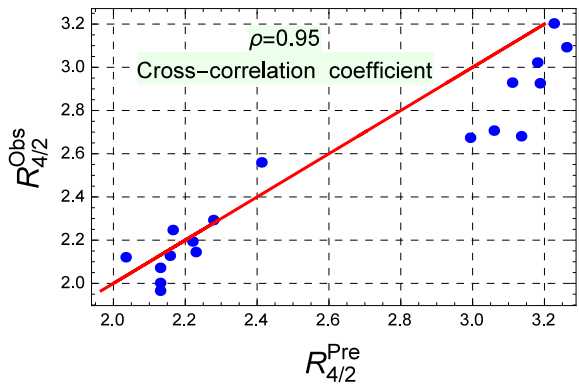}} 
		\caption{The correlation between the observed signature ratio $R_{4/2}^{Obs}$ and the predicted signature ratio $R_{4/2}^{Pred}$ .}
		\label{Fig_5}
	\end{figure}

\begin{figure}[H]
		\centering
		\rotatebox{0}{\includegraphics[height=65mm]{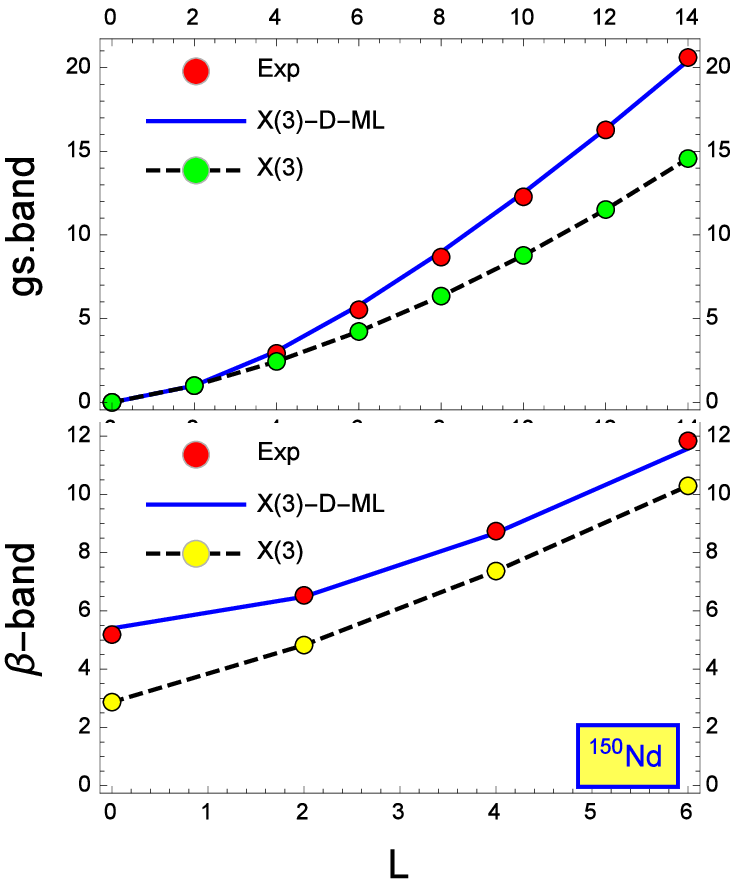}} 
		\rotatebox{0}{\includegraphics[height=65mm]{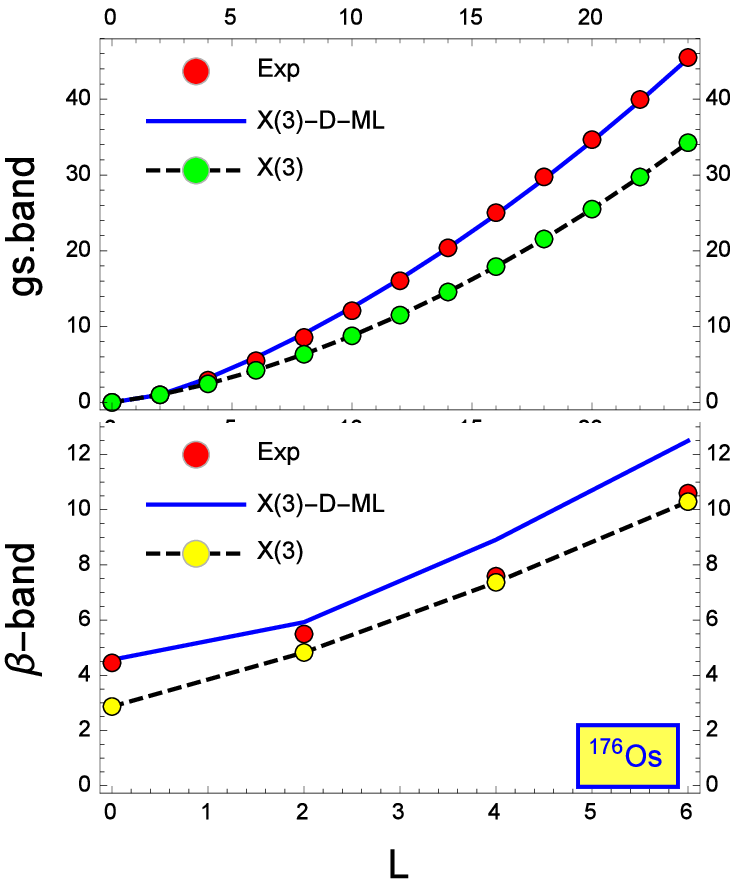}} 
		\rotatebox{0}{\includegraphics[height=65mm]{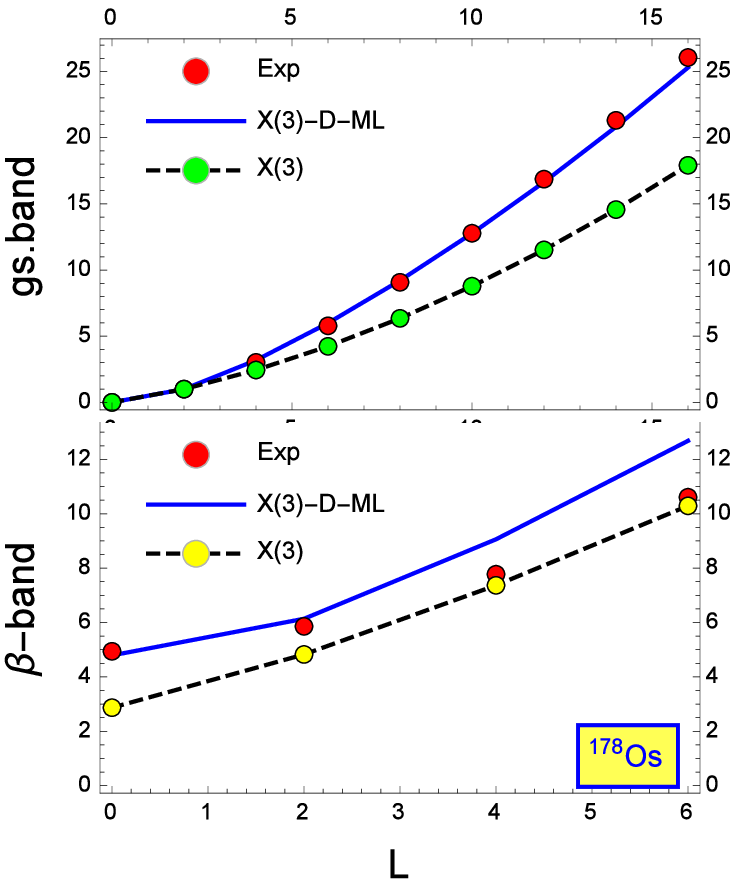}} 
		\rotatebox{0}{\includegraphics[height=65mm]{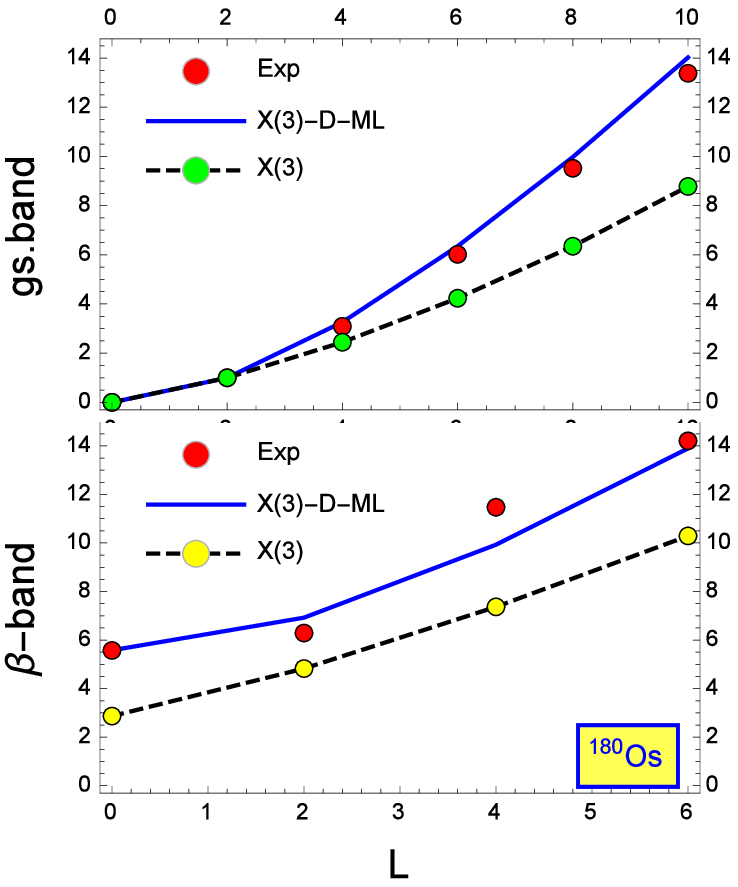}} 
		\rotatebox{0}{\includegraphics[height=65mm]{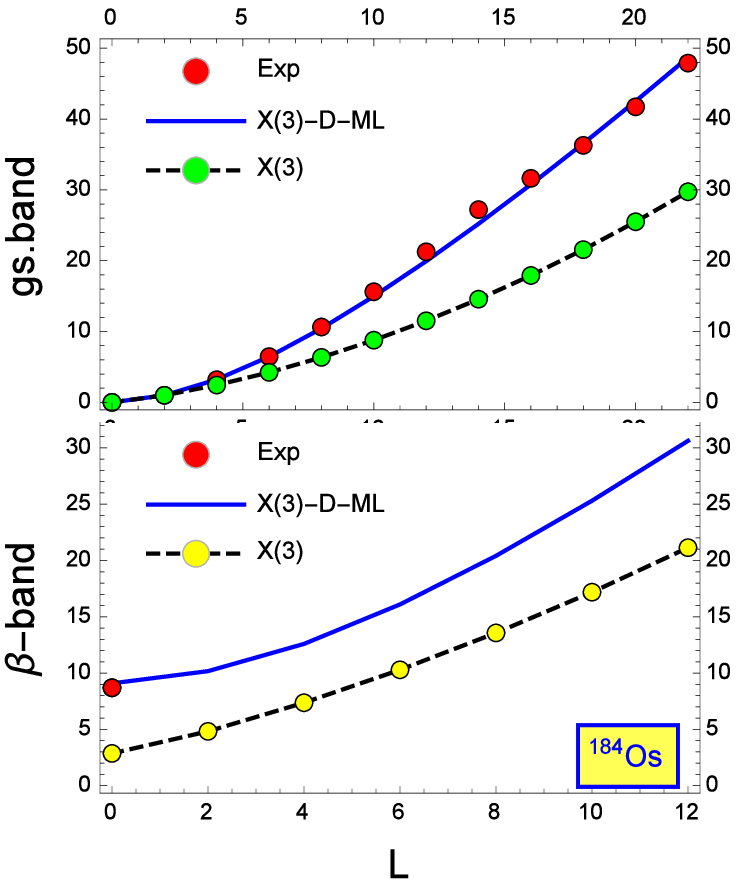}} 
		\rotatebox{0}{\includegraphics[height=65mm]{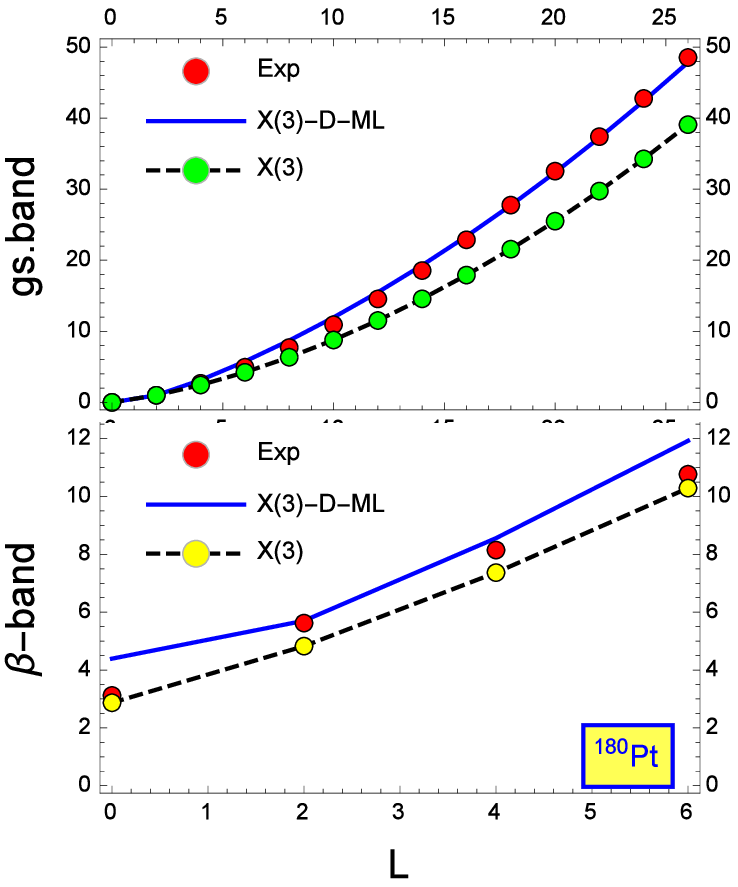}} 
		\rotatebox{0}{\includegraphics[height=65mm]{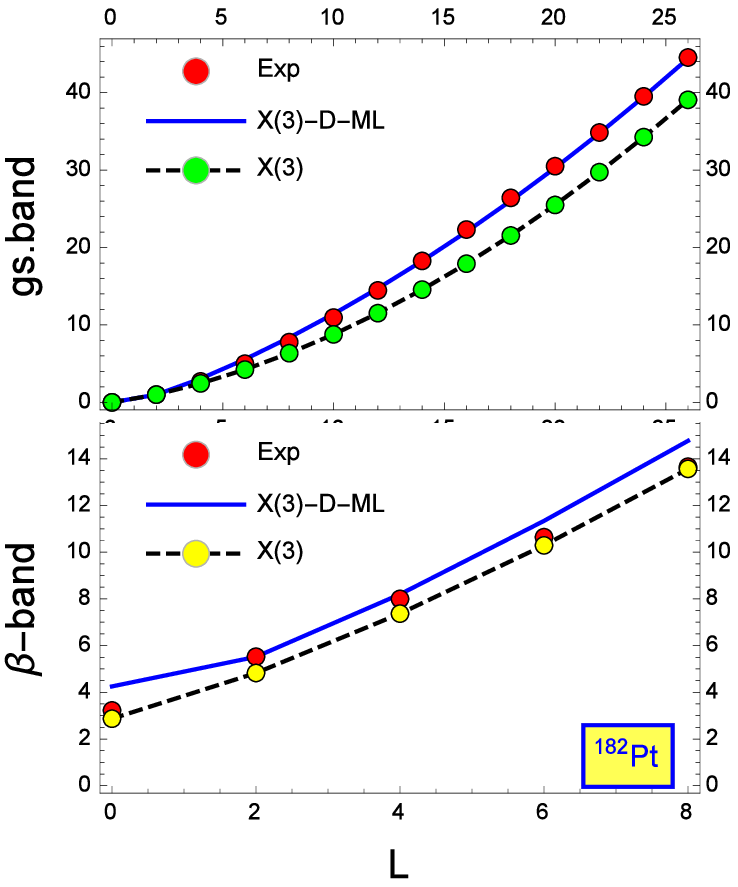}} 
		\rotatebox{0}{\includegraphics[height=65mm]{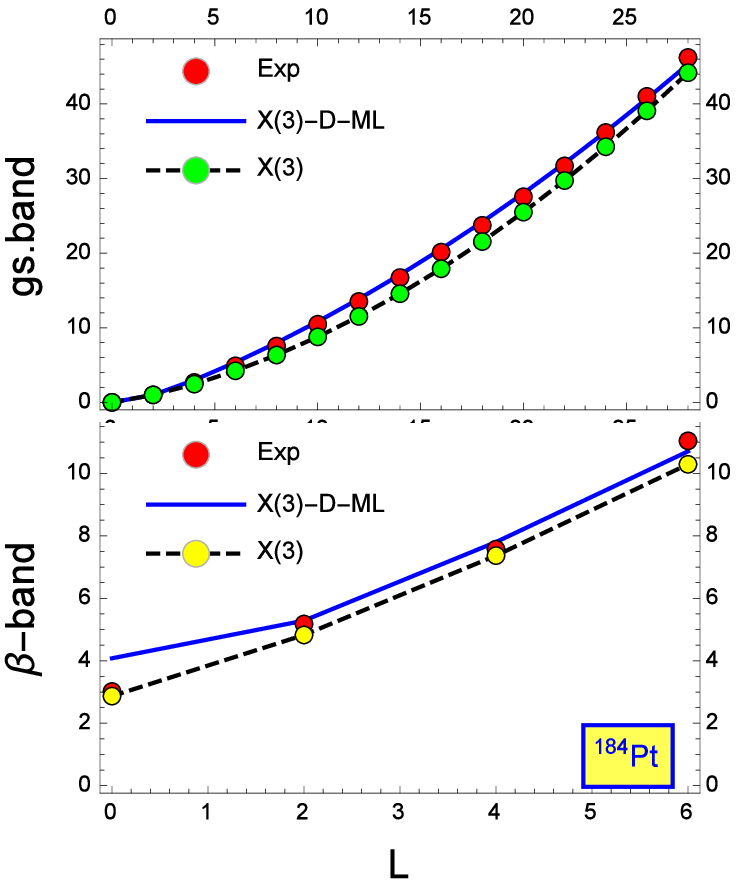}} 
		\rotatebox{0}{\includegraphics[height=65mm]{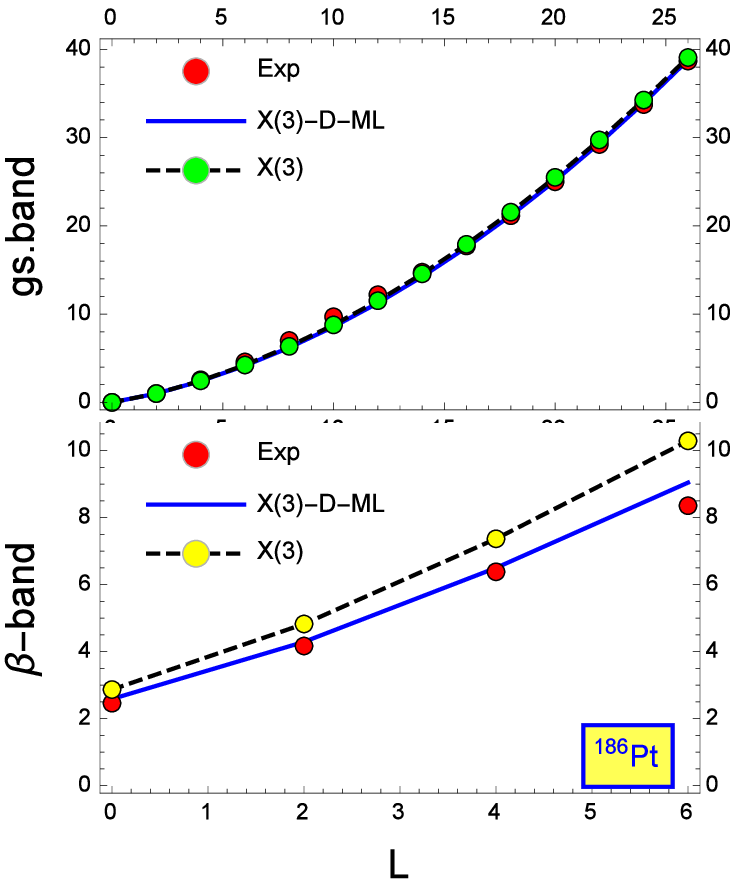}} 		
		\caption{Theoretical results for ground and first  excited $\beta$ bands energies normalized to the energy of the $2^+$ ground state are compared with the available experimental data\cite{b42} for $^{150}$Nd, $^{176-180}$Os, $^{184}$Os and $^{180-186}$Pt.  }
		\label{Fig_6}
	\end{figure}

\begin{figure}[H]
		\centering
		\rotatebox{0}{\includegraphics[height=65mm]{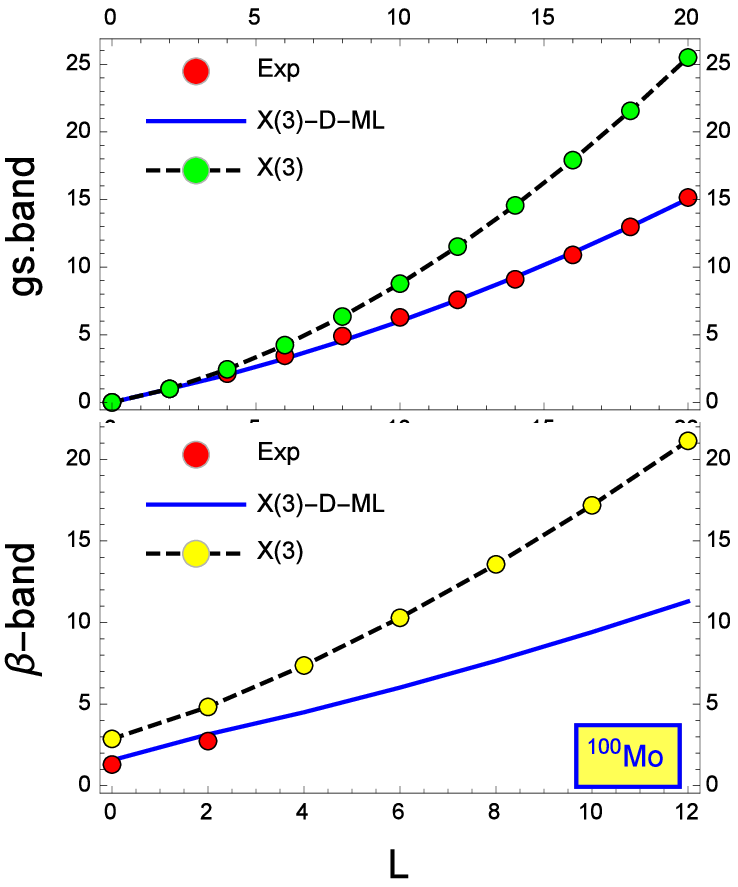}} 
		\rotatebox{0}{\includegraphics[height=65mm]{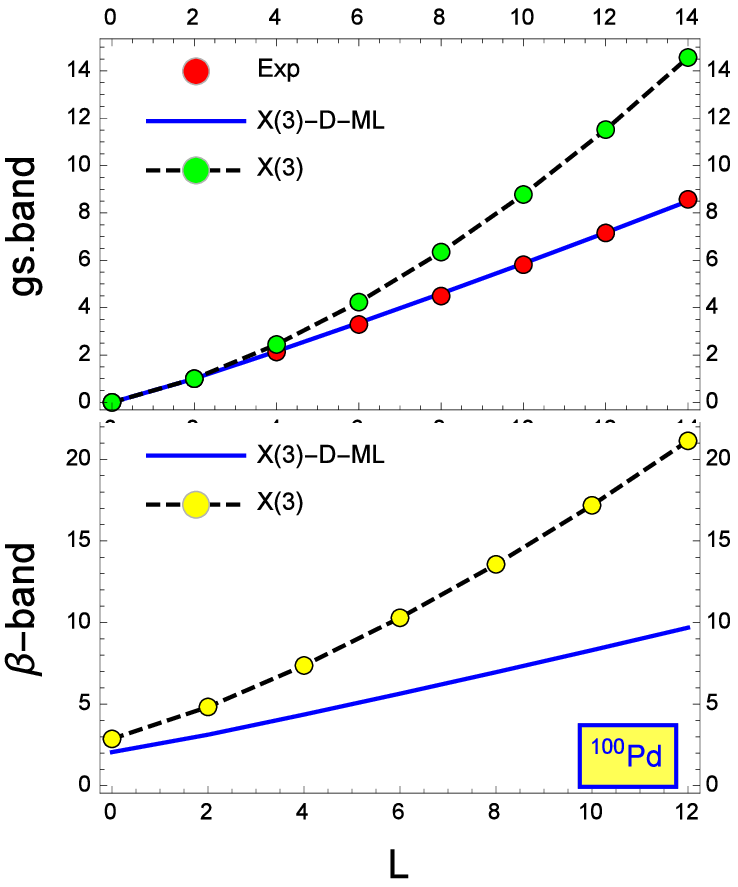}} 
		\rotatebox{0}{\includegraphics[height=65mm]{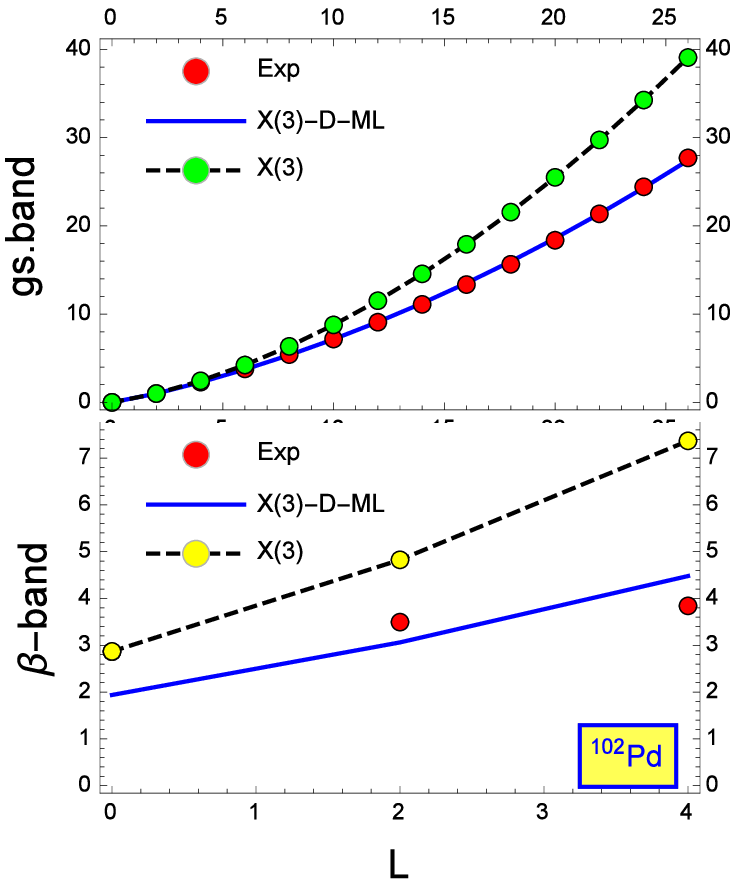}} 
		\rotatebox{0}{\includegraphics[height=65mm]{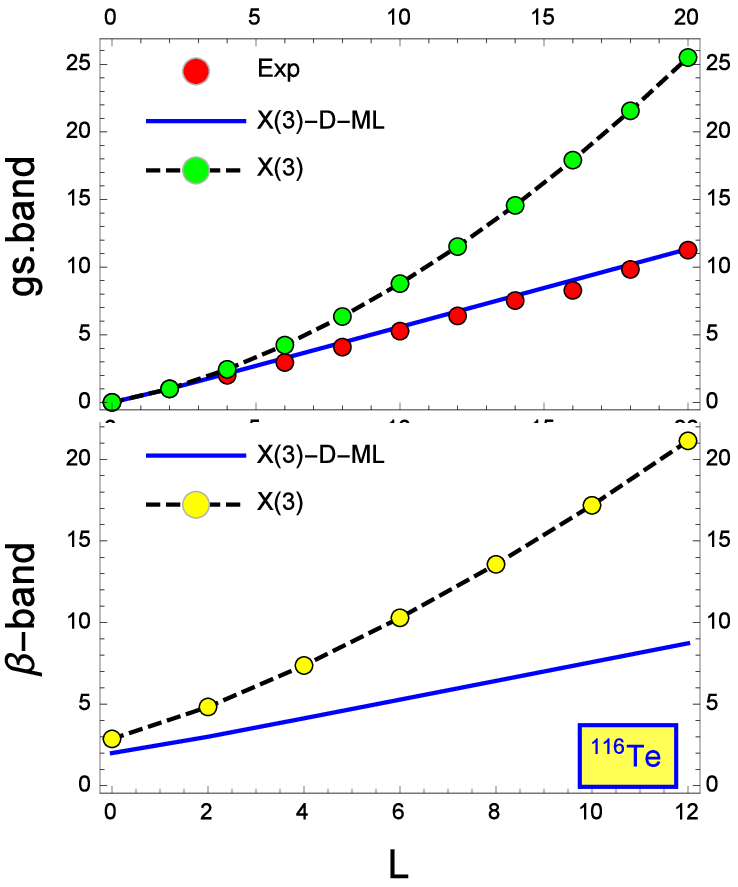}} 
		\rotatebox{0}{\includegraphics[height=65mm]{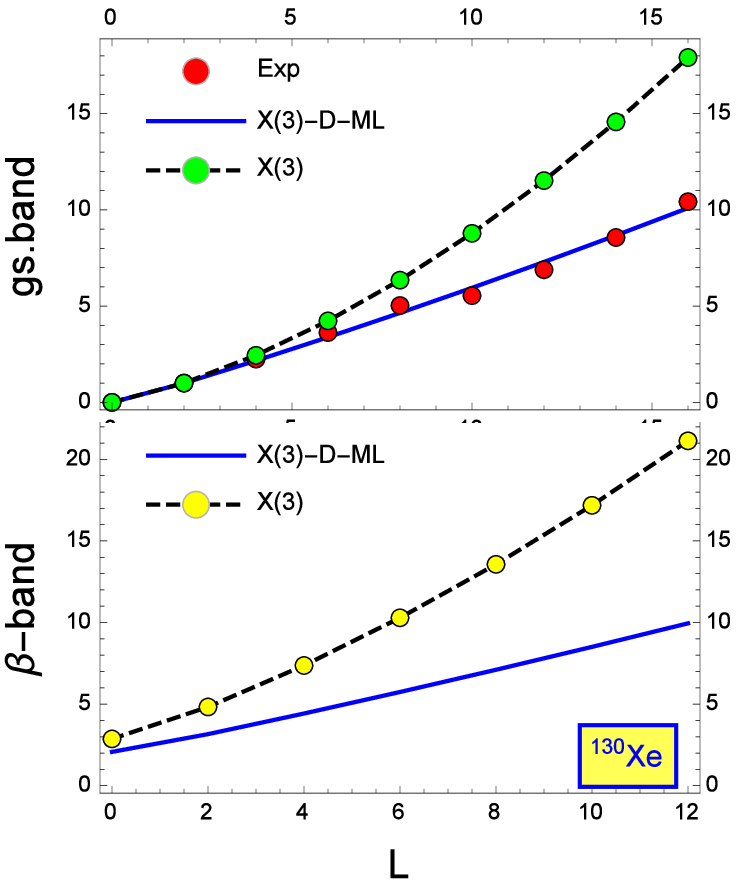}} 
		\rotatebox{0}{\includegraphics[height=65mm]{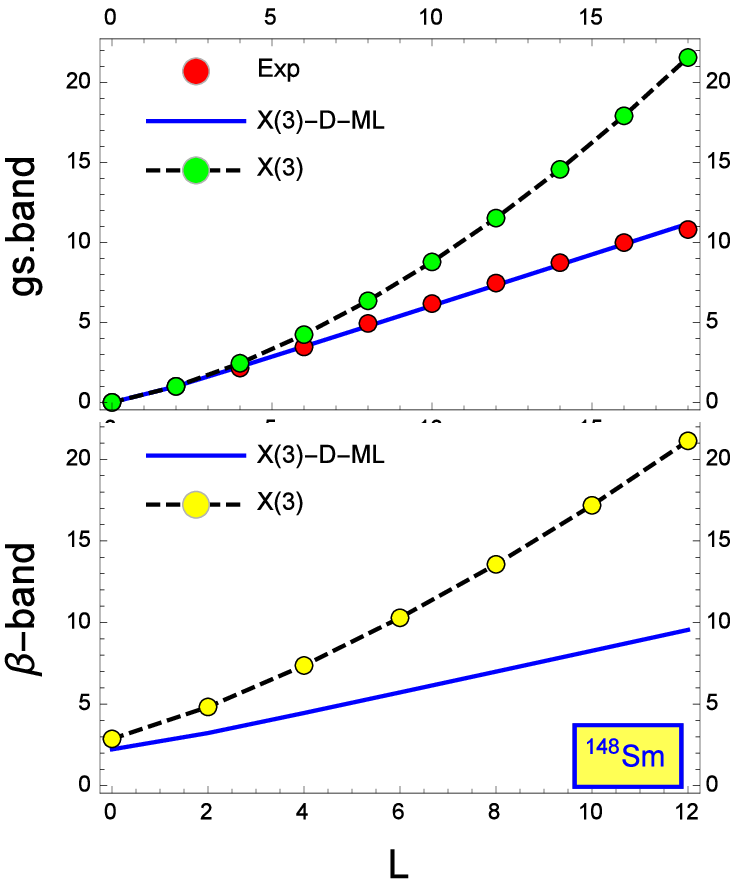}} 
		\rotatebox{0}{\includegraphics[height=65mm]{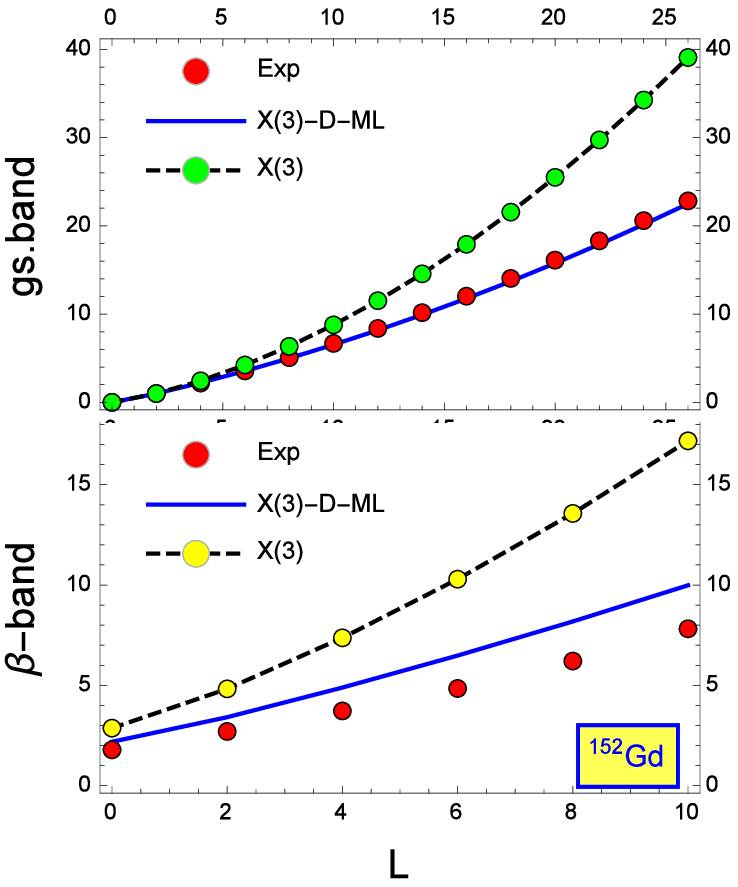}} 
		\rotatebox{0}{\includegraphics[height=65mm]{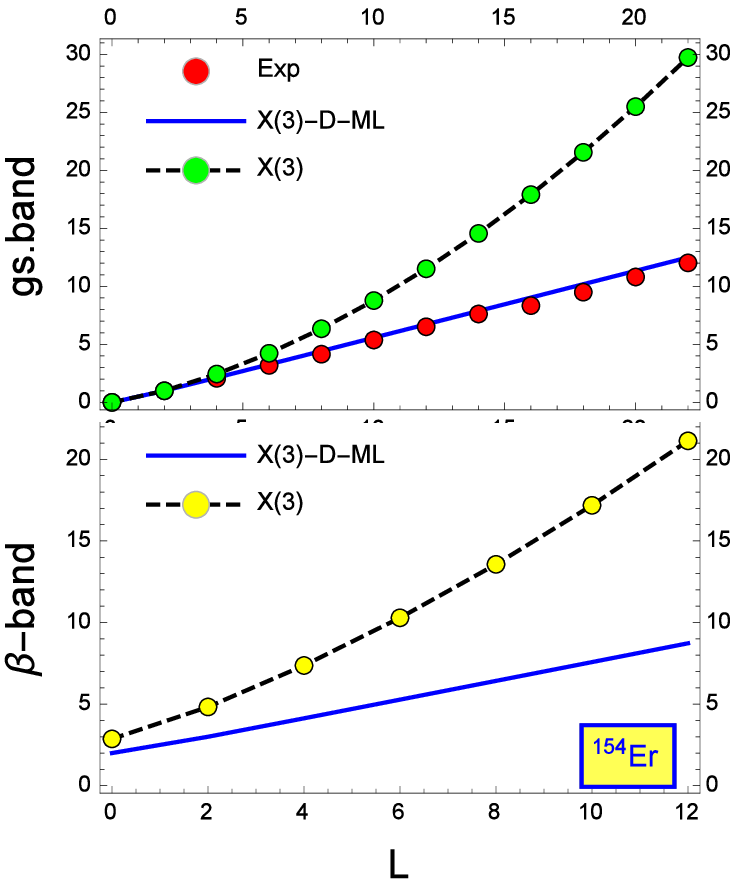}} 
		\rotatebox{0}{\includegraphics[height=65mm]{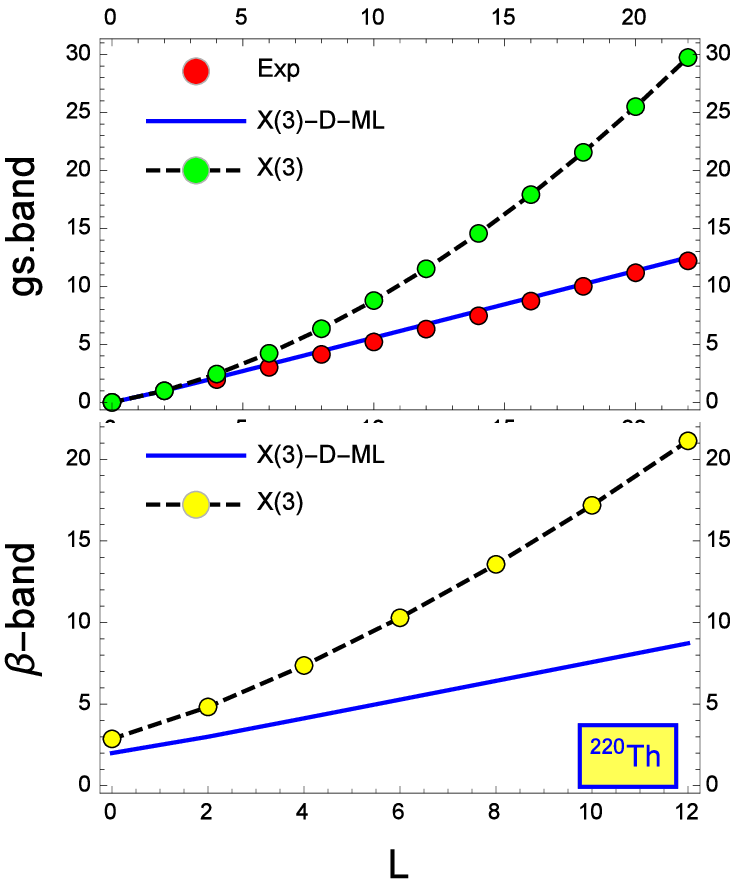}} 		
		\caption{Theoretical results for ground and first  excited $\beta$ bands energies normalized to the energy of the $2^+$ ground state are compared with the available experimental data\cite{b42} for $^{100}$Mo, $^{100}$Pd, $^{102}$Pd, $^{116}$Te, $^{130}$Xe, $^{148}$Sm, $^{152}$Gd, $^{154}$Er and $^{220}$Th.  }
		\label{Fig_7}
	\end{figure}

\begin{figure}[H]
	\centering	
	\rotatebox{0}{\includegraphics[height=47mm]{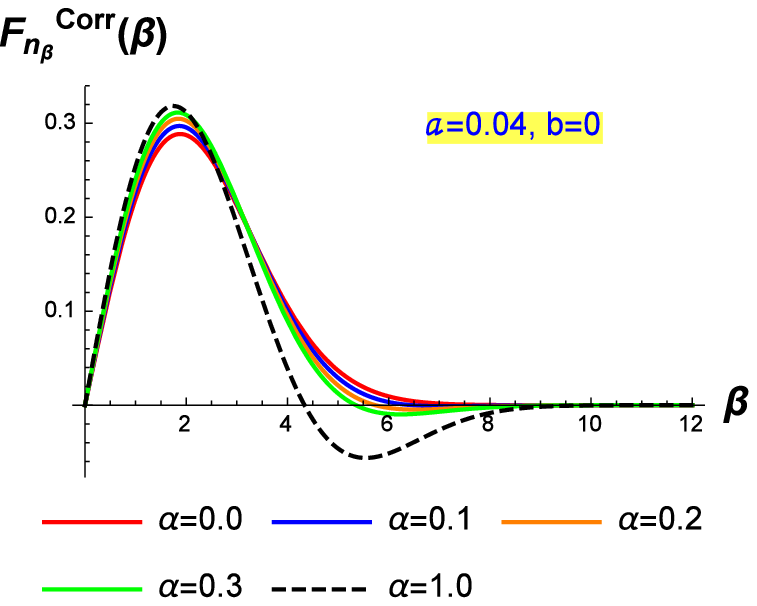}} 
	\rotatebox{0}{\includegraphics[height=47mm]{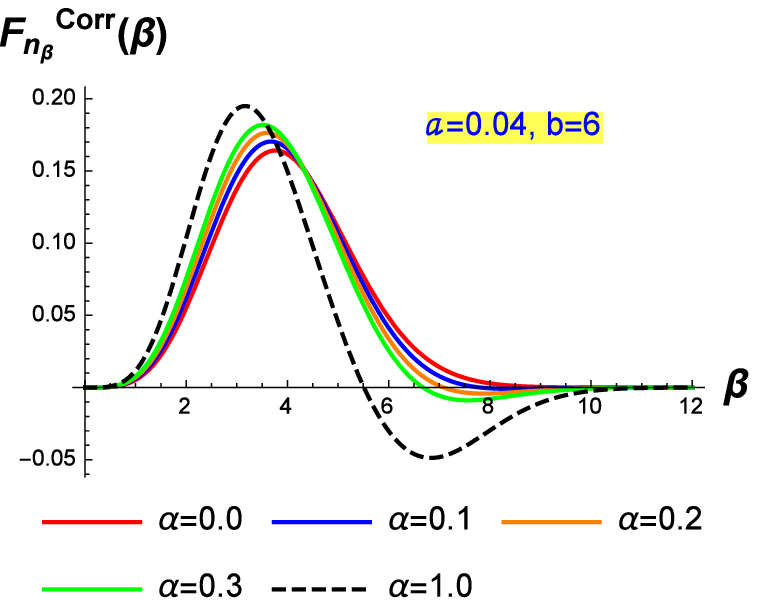}}
	\rotatebox{0}{\includegraphics[height=47mm]{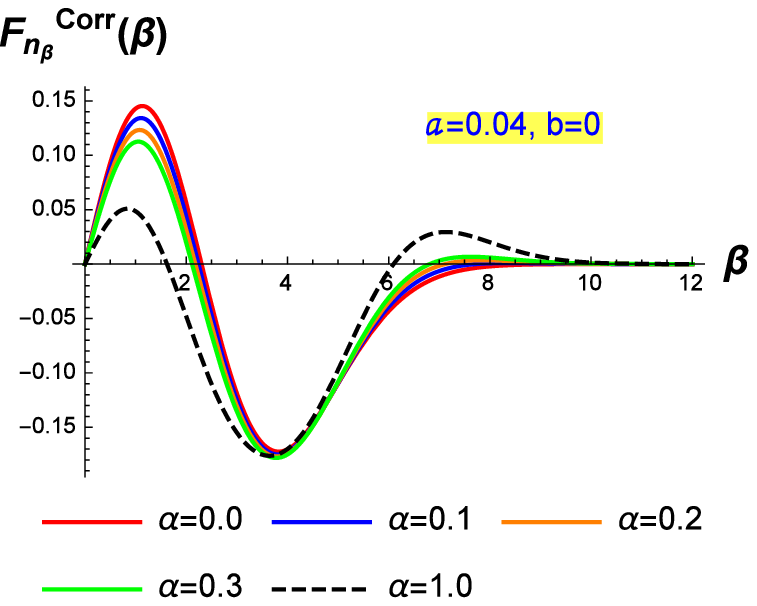}} 
	\rotatebox{0}{\includegraphics[height=47mm]{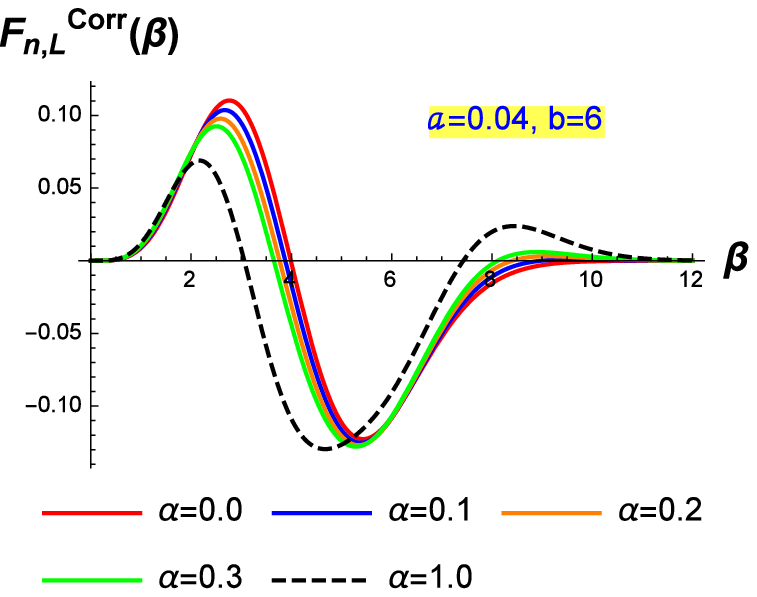}}
	\caption{The corrected wave function drawn as a function of $\beta$ and $\alpha$ for ground state (Upper panel) and for the first excited $\beta$ state (Lower panel).}
	\label{Fig_f1}
\end{figure}

\begin{figure}[H]
	\centering	
	\rotatebox{0}{\includegraphics[height=47mm]{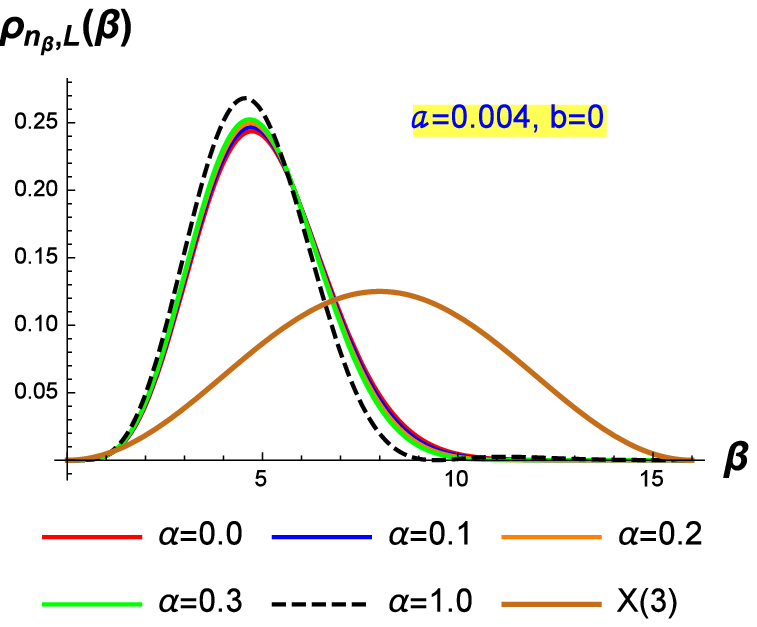}} 
	\rotatebox{0}{\includegraphics[height=47mm]{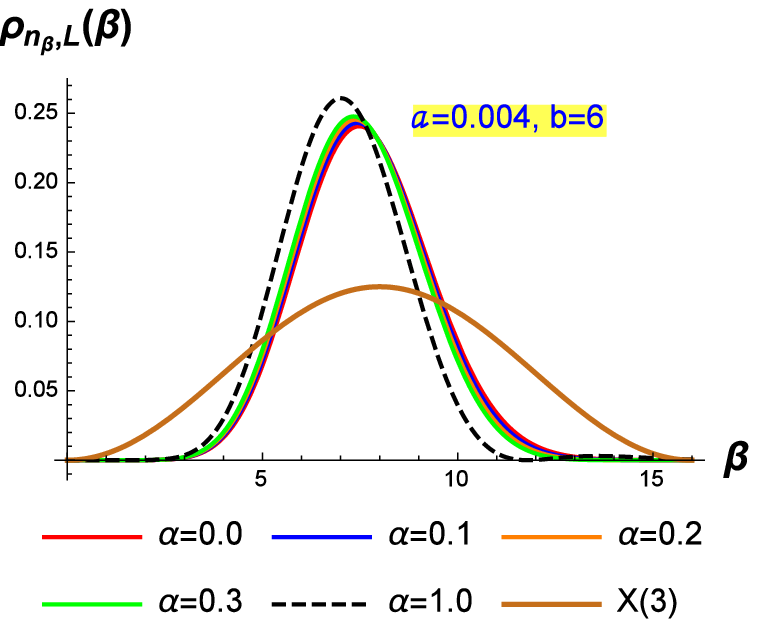}}
	\rotatebox{0}{\includegraphics[height=47mm]{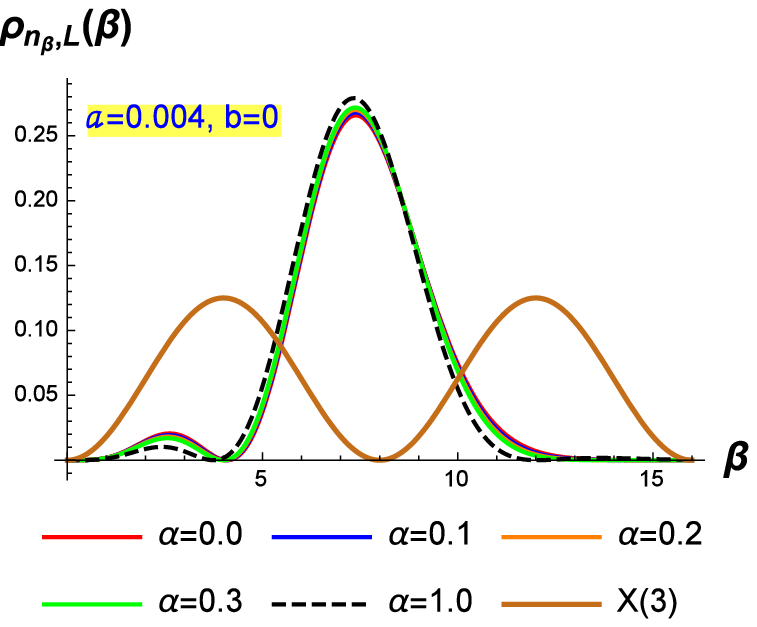}} 
	\rotatebox{0}{\includegraphics[height=47mm]{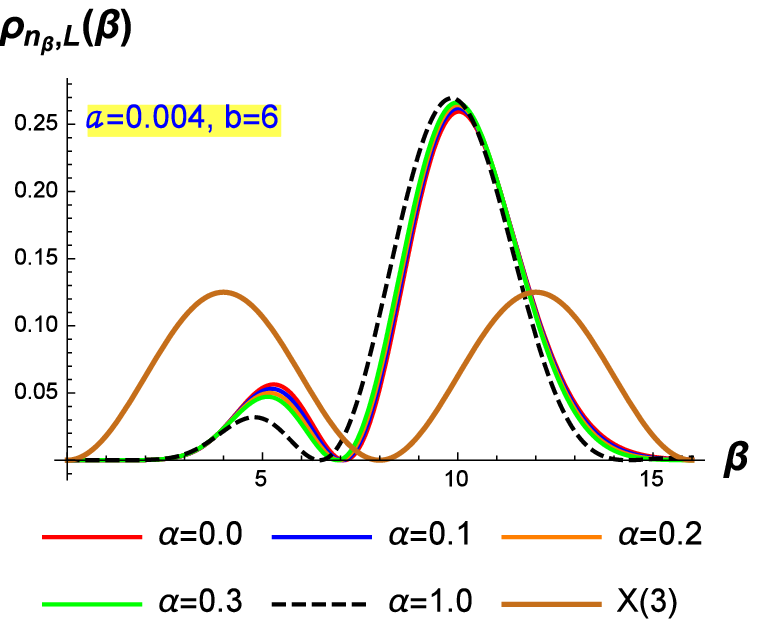}}
	\caption{The density of probability distribution as a function of $\beta$ and $\alpha$ for ground state (Upper panel) and for the first excited $\beta$ state (Lower panel). The profiles of the probability distribution of X(3) model is also shown for comparison.}
	\label{Fig_f2}
\end{figure}

	The model established in this work, called X(3)-D-ML, is adequate for the description of $\gamma$-rigid nuclei for which the $\gamma$ parameter is fixed to $\gamma=0$. Basically, the energy levels of the ground state band as well as of the first $\beta$ vibrational band are  characterized by the principal quantum number $n_{\beta}=0$ and $n_{\beta}=1$, respectively. Besides, no $\gamma$-bands appear in the present model as expected, because the $\gamma$-degree of freedom has been initially frozen to $\gamma=0$.
	It is immediate to see that the energy spectrum of our model has three adjustable parameters, namely : the minimal length parameter $\alpha$, and the two parameters of Davidson potential: $a$ and $b$ . In the axially symmetric case, the shape phases of nuclei are parametrized by a nuclear deformation parameter which is not an immediately measurable observable. Then, instead one usually describes these shape phases as function of the ratio $R_{4/2}$ between the lowest two collective energy levels $4_g^+$ and $2_g^+$. It is interesting to see that at each shape phase corresponds a dynamical symmetry whose signature is a specific value of $R_{4/2}$ as in IBM approach\cite{b9}. Therefore, it should be useful to examine the applicability of our model by investigating the dependence on the free parameters of the energy spectrum and then the signature ratio $R_{4/2}$. Moreover, the evolution of the energy spectrum normalized to the first excited state for the ground band (left) and the first $\beta$ band (right) as a function of angular momentum L is depicted in Fig. \ref{Fig_1} and Fig. \ref{Fig_2}, for different values of the parameter $\alpha$ ranging from $0$ to $1$ which enclose the existence region of the model. 
	By analyzing Fig. \ref{Fig_1}, it can be observed that the region allowed by our model for both the ground and the $\beta$ bands is located below X(3) model  for the parameters $a=0.004$ and $b=0$. On the other hand, in Fig. \ref{Fig_2}, the region allowed by our model is located in this case above X(3) model for the parameters $a=0.004$ and $b=6$. To clarify these remarks, we plot in figure \ref{Fig_3} and \ref{Fig_4} the deviation, between the energy spectrum of our model and the energy of X(3) model, defined by the quantity $D_{X(3)}^{D-ML}=E_{n_{\beta},L}^{D-ML}/E_{n_{\beta},L}^{X(3)}$, which encloses also the existence region of the model, as functions of L for arbitrary values of the parameter $a$ and $b$. These theoretical results clearly show the flexibility of our model for correctly describing the structural properties of nuclei. Now, it's desirable to place the present results in a wider picture of other similar collective models, which is done in Table \ref{Tab_1}, where the shapes  of the $\beta$ potential  are presented.
	\begin{table}[!h]
		\caption{The potentials in the $\beta$ variable for the relevant $\gamma$-rigid solutions.}
		\begin{center}
			\begin{tabular}{|c|c|}
				\hline
				Models&$\beta$ shape potential\\
				\hline
				X(3)\cite{b5}&0,$\;$if$\;$$\beta\leq\beta_{\omega}$,$\;$\\
				&$\infty$,$\;$if$\;$$\beta>\beta_{\omega}$\\
				X(3)-$\beta^{2}$\cite{b23}&$\sim\beta^{2}$\\
				X(3)-$\beta^{4}$\cite{b23}&$\sim\beta^{4}$\\
				X(3)-$\beta^{6}$\cite{b24}&$\sim\beta^{6}$\\
				QAOP\cite{b23}&$\frac{1}{2}\alpha_{1}\beta^{2}+\alpha_{2}\beta^{4},$\\
				&$\alpha_{1}\geq0,\;\alpha_{2}>0$\\
				SAOP\cite{b24}&$\frac{1}{2}\alpha_{1}\beta^{2}+\alpha_{2}\beta^{6},$\\
				&$\alpha_{1}\geq0,\;\alpha_{2}>0$\\
				X(3)-Sextic\cite{b25,b33}&$(b^{2}-4ac)\beta^{2}+2ab\beta^{4}+a^{2}\beta^{6}$,\\
				&$c,a>0,\;b\in R$\\
				X(3)-ML\cite{b15}& 0,$\;$if$\;$$\beta\leq\beta_{\omega}$,$\;$\\
				&$\infty$,$\;$if$\;$$\beta>\beta_{\omega}$\\
				Present (X(3)-D-ML)& $a\beta^{2}+\frac{b}{\beta^2}$\\
				&$a>0, \ b\geq0,\ \beta_0=(b/a)^{1/4}$\\
				\hline
			\end{tabular}
		\end{center}
		\label{Tab_1}
	\end{table}	
	 Moreover, from figures \ref{Fig_1} and \ref{Fig_2}, one can see that, for different arbitrary values of scaling parameters $a$ and $b$, the centrifugal potential part in Davidson potential plays a crucial role in defining the allowed region by our model, which becomes larger in respect to the pure Harmonic Oscillator case. Also, in the absence of the centrifugal potential part ($b=0$), our model X(3)-D-ML tends to the X(3) one for $\alpha\rightarrow 1$, while in its presence ($b\neq0$), the situation is inverted. In addition, the effect of minimal length increases more, as a function of angular momentum, in the presence of centrifugal potential part comparatively to the pure Harmonic Oscillator. Such a situation is well illustrated in Figures \ref{Fig_3} and \ref{Fig_4}.
	  Besides, from these figures, it is apparent that the gap between our model and the X(3) one increases further for higher angular momentum states. Such a fact will have a positive effect in compensation of the defect of the model without minimal length ($\alpha=0$) when reproducing the experimental data as will be seen afterwards. As mentioned above, the introduction of scaling parameters in Davidson potential has been done with the aim of getting physical values for the minimum of the potential ($\beta_0=\left(\frac{b}{a}\right)^{1/4}<1$) as can be seen from Table \ref{Tab_2}. The same approach has been performed  in Ref\cite{b36_ad}, in the framework of the Kratzer potential. In earlier works \cite{b19,b22,b38,b39,b40}, this minimum was problematic since its obtained values were unphysical ($\beta_0>1$) in respect to the nuclear deformation. In the same table ( Table \ref{Tab_2}), we present the bandhead ratios $R_{4/2}$ calculated by our model compared to the experimental data. The agreement between theory an experiment is evaluated by the r.m.s deviation $\sigma$, which takes values lower than unit for all studied nuclei. Such an agreement is corroborated by the obtained cross-correlation coefficient $\rho=0.95$ (Fig. \ref{Fig_5}). 
	However, among all studied nuclei, the isotopes ${}^{100}$Mo, ${}^{100-102}$Pd, ${}^{116}$Te, ${}^{130}$Xe, ${}^{148}$Sm, ${}^{150}$Nd, ${}^{152}$Gd, ${}^{154}$Er, ${}^{176-180}$Os, ${}^{184}$Os, ${}^{180-186}$Pt and ${}^{220}$Th seem to be the best candidates for our model. Such a result is well illustrated in figures \ref{Fig_6} and \ref{Fig_7}. From these figures, one can see that, for these good candidate nuclei, there is an excellent agreement between theory and experiment in both ground state and the first $\beta$-bands. It should be noted that for isotopes $^{100}$Pd, $^{116}$Te, $^{130}$Xe, $^{148}$Sm, $^{154}$Er and $^{220}$Th, the experimental data for the first $\beta$-band are not available, so only theoretical predictions are presented. Moreover, from these figures, we can observe the effect of the minimal length, which becomes conspicuous in high momentum states. In addition, by comparing our results in Fig. \ref{Fig_6} with those plotted in Fig. 7 of Ref. \cite{b23}, which have been obtained with a quartic anharmonic oscillator potential, it appears clearly that,  thanks to the effect of minimal length, our results become better with the increase of angular momentum. Such a fact is coherent with the quintessence of the minimal length concept as introduced in string theory\cite{b26,b27,b28}. Besides, from Table \ref{Tab_2}, we observe that the nuclei, which are above the isotope ${}^{186}$Pt, behave as a rotor ($b\neq 0$). Their bandhead ratio is $R_{4/2}\succeq 3$. On the other hand, the nuclei below ${}^{186}$Pt behave as a vibrator ($b\rightarrow 0$). Their bandhead ratio is $R_{4/2}\prec 2.4$.
		\begin{table}[!h]
		\caption{The parameters obtained from the fits visualized in Figures \ref{Fig_6} and \ref{Fig_7}.}
		\label{Tab_2}
		\resizebox{0.88\textwidth}{!}{\begin{minipage}{\textwidth}
				\begin{tabular}{c|c|c|c|c|c|c|c|}
					\cline{2-8}
					& \multicolumn{2}{c|}{Signature ratios} & \multicolumn{4}{c|}{Parameters}  & The mean deviation \\ \hline
					\multicolumn{1}{|l|}{Nucleus}  & R$_{4/2}$(Observed) & R$_{4/2}$(Predicted)&  $a$    &   $b$   &  $\alpha$ & $\beta_0$ & $\sigma$    \\ \hline
					\multicolumn{1}{|c|}{$^{150}$Nd}& 2.929              & 3.110               & 13.662  & 2.1339  & 0.0072    & 0.6286    & 0.2652  \\ \hline
					\multicolumn{1}{|c|}{$^{176}$Os}& 2.926              & 3.187               & 12.472  & 1.0202  & 0.0155    & 0.5347    & 0.6202  \\ \hline
					\multicolumn{1}{|c|}{$^{178}$Os}& 3.022	             & 3.180               & 11.730  & 1.4253  & 0.0137    & 0.5904    & 0.7341    \\ \hline
					\multicolumn{1}{|c|}{$^{180}$Os}& 3.093            	 & 3.261               & 4.5109  & 2.3798  & 0.0199    & 0.8522    & 0.6031    \\ \hline
					\multicolumn{1}{|c|}{$^{184}$Os}& 3.203              & 3.226               & 9.0008  & 8.9119  & 0.0028    & 0.9975    & 0.8066      \\ \hline
					\multicolumn{1}{|c|}{$^{180}$Pt}& 2.681              & 3.135               & 5.9160  & 0.8601  & 0.0203    & 0.6175    & 0.6824  \\ \hline
					\multicolumn{1}{|c|}{$^{182}$Pt}& 2.707              & 3.059               & 10.141  & 0.8618  & 0.0129    & 0.5399    & 0.4678        \\ \hline
					\multicolumn{1}{|c|}{$^{184}$Pt}& 2.674            	 & 3.000               & 7.5606  & 0.7397  & 0.0126    & 0.5593    & 0.4515        \\ \hline
					\multicolumn{1}{|c|}{$^{186}$Pt}& 2.560            	 & 2.412               & 0.9662  & 0.0000  & 0.2045    & 0.0000    & 0.4659          \\ \hline
					\multicolumn{1}{|c|}{$^{100}$Mo}& 2.121              & 2.034               & 3.1127  & 0.3066  & 0.0279    & 0.5602    & 0.2084           \\ \hline
					\multicolumn{1}{|c|}{$^{100}$Pd}& 2.128            	 & 2.157               & 3.0368  & 0.0000  & 0.0049    & 0.0000    & 0.0568            \\ \hline
					\multicolumn{1}{|c|}{$^{102}$Pd}& 2.293              & 2.278               & 4.5414  & 0.0000  & 0.0251    & 0.0000    & 0.3199             \\ \hline	
					\multicolumn{1}{|c|}{$^{116}$Te}& 2.002              & 2.130               & 8.5729  & 0.0000  & 0.0000    & 0.0000    & 0.3408              \\ \hline
					\multicolumn{1}{|c|}{$^{130}$Xe}& 2.247              & 2.165               & 0.0394  & 0.0000  & 0.0560    & 0.0000    & 0.2705               \\ \hline
					\multicolumn{1}{|c|}{$^{148}$Sm}& 2.145            	 & 2.229               & 4.7919  & 0.0957  & 0.0000    & 0.3759    & 0.1524                \\ \hline
					\multicolumn{1}{|c|}{$^{152}$Gd}& 2.194              & 2.220               & 0.3495  & 0.0000  & 0.0555    & 0.0000    & 0.8421                 \\ \hline
					\multicolumn{1}{|c|}{$^{154}$Er}& 2.072            	 & 2.130               & 16.936  & 0.0000  & 0.0000    & 0.0000    & 0.3755                 \\ \hline
					\multicolumn{1}{|c|}{$^{220}$Th}& 1.966              & 2.130               & 2.0056  & 0.0000  & 0.0000    & 0.0000    & 0.2728                  \\ \hline
				\end{tabular}
		\end{minipage}}
	\end{table}	   
	 Moreover, from figures \ref{Fig_6} and \ref{Fig_7}, one can observe similar behaviours to those depicted in Figures \ref{Fig_1} and \ref{Fig_2} for energy ratio in ground and first $\beta$-bands in relation with the values of the parameters $b$ and $\alpha$. Such a fact reveals the existence of some correlation between  minimal length and centrifugal potential as was outlined above  from Fig. \ref{Fig_2}. Moreover, from Fig. \ref{Fig_6}, one can see that  the obtained ratios by our model, for shape rotor nuclei, are localised above  the critical point symmetry $X(3)$, while those obtained for shape vibrator nuclei (Fig. \ref{Fig_7}) are below this critical point in concordance with what was mentioned above in relation with the centrifugal potential. However, in the case of the isotope ${}^{186}$Pt, which is at the middle position between the remainder nuclei, both models $X(3)$-D-ML and $X(3)$ coincide perfectly with the experimental data in the ground state band (Fig. \ref{Fig_6}), while in the first $\beta$-band, our model X(3)-D-ML coincides with the experiment and the gap with the X(3) model is small in comparaison with the other studied nuclei. In addition, the obtained model parameters and the band head ratio $R_{4/2}$ are particular, namely: $a\simeq1$, $b=0$, $\alpha=0.2$ (the greatest value) and $2.4<R_{4/2}<3$. This isotope corresponds to a critical point between vibrator and rotor behaviours in the studied transitional nuclei. On the other hand, the physical meaning of parameter $\alpha$ in the frame of collective excitations, can be explained by studying the corrected wave function as well as the corresponding density probability distribution. The behaviours of these two quantities versus $\beta$ for arbitrary values of the Davidson potential parameters $a$ and $b$ and for $\alpha$ values varying from $0$ to $1$  are depicted in figures \ref{Fig_f1} and \ref{Fig_f2} for the ground $0_g^+$state and the excited  $0_{\beta}^+$ state, respectively.  From  figure \ref{Fig_f1}, one can see that for values of $\alpha$ up to $0.3$, the behaviour of the wave function is regular corresponding well and truly to the ground state. However, beyond this limit value, the wave function starts showing a nodal point which means that we are getting away from the ground state . The same phenomenon occurs in the excited $0_{\beta}^+$ state, especially when the centrifugal term is present. Therefore, this observation shows that the minimal length parameter $\alpha$ has a physical boundary as is clearly reflected in concrete nuclei from Table \ref{Tab_2}. Moreover, from figure \ref{Fig_f2}, one can see that, in the case of a pure Harmonic Oscillator  $(b=0)$, the density probability peaks are shifted backward on the left side in respect to the X(3) peak, while in the presence of Davidson centrifugal potential part $(b\ne0)$, they are shifted forward on the right side in concordance with the presented situation in figures 1 and 2. We conclude our discussion by stressing that the calculations of electromagnetic transition rates (especially the Monopole transition probability E0) can be obtained from the corrected wave functions(\ref{E28_a}), following the same approach presented in Ref.\cite{b41}. Although, the calculation of this quantity, in the framework of our model, is a bit complicated but it remains a very important topic for discussion which will be the subject of the next work. \textit{In fine}, we have to notice that the treatment of the present problem within AIM or any other analytical method is mathematically complicated. So, the use of QPM, for the first time within this thematic, allowed us to overcome such a difficulty and therefore will pave the way for further easy applications of the minimal length formalism.
	\newpage
\section{Conclusions}
	In our pioneering work\cite{b15}, we have introduced the minimal length concept in nuclear structure through Bohr- Mottelson model where we got an improved version of the model X(3) called X(3)-ML. However, the solution of the radial equation (in $\beta$) of X(3)-ML for a potential other than the square well is  mathematically more complicated issue. In order to overcome such a difficulty, in the present paper we used, for the first time, a quantum perturbation method. Therefore, closed-form analytical formula for the energy of the ground and the $\beta$ bands was derived for prolate $\gamma$-rigid nuclei within Davidson potential. Moreover, in this study, a correlation between minimal length and the centrifugal part of this potential has been revealed making the new elaborated model X(3)-D-ML more suitable  for describing the properties  of nuclei having a structure at or close to the X(3) limit. In addition, in order to get physical values for the minimum of Davidson potential in respect to the nuclear deformation, unlike many other previous works, we introduced two scaling parameters. Besides, the use of a quantum perturbation method within this thematic will allow to tackle the mathematical problem related with the utilization of the minimal length formalism in further applications with other sophisticated potentials. 
	\section{Appendix A}
	In this Appendix we present the calculations of the explicit expressions of mean values $\overline{\beta^{t}}(t=2,4,-2,-4)$. So, we have:
	\begin{align}
	\overline{\beta^{t}}=&\langle\psi_{n_{\beta},L}^{(0)}\left\vert \beta^t\right\vert \psi_{n_{\beta},L}^{(0)}\rangle\\
	=&\int_{0}^{\infty}\psi_{n_{\beta},L}^{(0)}(\beta)\beta^t\psi_{n_{\beta},L}^{(0)}(\beta)\beta^2d\beta\\
	=& N_{n_{\beta},L}^2\int_{0}^{\infty}\beta^{(t+2p)}e^{-2q\beta^2}\left[\mathcal{L}_{n_{\beta}}^{(p-\frac{1}{2})}\left(2q\beta^2\right)\right]^2d\beta
	\label{E29}
	\end{align}
	By using a new variable $s=2q\beta^2$, the above integral leads to, 
	\begin{align}
	\overline{\beta^{t}}= C_{n_{\beta},L}\int_{0}^{\infty}s^{(\frac{t}{2}+p-\frac{1}{2})}e^{-s}\left[\mathcal{L}_{n_{\beta}}^{(p-\frac{1}{2})}\left(s\right)\right]^2ds
	\label{E30}
	\end{align}
	with,
	\begin{equation}
	C_{n_{\beta},L}=\frac{N_{n_{\beta},L}^2\Gamma(1+n_{\beta})^2\Gamma(p+\frac{1}{2})^2}{2^{\left(\frac{t}{2}+p+\frac{3}{2}\right)}q^{\left(\frac{t}{2}+p+\frac{1}{2}\right)}\Gamma(n_{\beta}+p+\frac{1}{2})^2}
	\label{E31}
	\end{equation}
	In order to obtain the normalization factor and matrix elements of some physical functions we have to derive the exact expressions of (\ref{E30}).
	In the case of $t= 0$, the integral is easily obtained via the following formula:
	\begin{equation}
	\int_{0}^{\infty}s^{k}e^{-s}\left[\mathcal{L}_{n_{\beta}}^{k}\left(s\right)\right]^2ds=\frac{\Gamma(n_{\beta}+k+1)}{n_{\beta}+1}
	\label{E32}
	\end{equation}
	According to this we find,
	\begin{align}
	N_{n_{\beta},L}=\left[\frac{2^{\left(p+\frac{3}{2}\right)}q^{\left(p+\frac{1}{2}\right)}\Gamma(n_{\beta}+p+\frac{1}{2})}{\Gamma(n_{\beta}+1)\Gamma(p+\frac{1}{2})}\right]^{\frac{1}{2}}
	\label{E33}
	\end{align}
	For the case of $t\ne0$, the mean values can be calculated by using the generalized formula (BU 142(19)) in \cite{b37}. Thus, we obtain the following equation,
	\begin{equation}
	\overline{\beta^{t}}=C_{n_{\beta},L}\cdot\frac{\Gamma(p+\frac{t}{2}+\frac{1}{2})\Gamma(n_{\beta}+p+\frac{1}{2})}{n_{\beta}!^2\Gamma(p+\frac{1}{2})}\cdot S_{n_{\beta}}^{(p,q)}
	\label{E34}
	\end{equation}
	with,
	\begin{align}
	S_{n_{\beta}}^{(p,q)}=\Bigg\{\frac{d^{n_{\beta}}}{dh^{n_{\beta}}}\left[\frac{{}_2 \mathcal{F}_1\left(\frac{1+u+v}{2},\frac{2+u+v}{2};u+1;\mu^2/\chi^2\right)}{\left(1-h\right)^{1+u}\chi^{1+u+v}}\right] \Bigg\}_{h=0}
	\label{E35}
	\end{align}
	where, we have used the following parametrization:
	\begin{equation}
	u=p-\frac{1}{2},\ v=\frac{t}{2},\ \mu^2=\frac{4h}{(1-h)^2},\ \chi=\frac{1+h}{1-h}
	\label{E36}
	\end{equation}
	from which we may derive the analytical results of the mean values by making use of the mathematical symbolic computation programs like Mathematica or Maple. Therefore, the analytical expressions of the mean values $\overline{\beta^{t}}(t=2,4,-2,-4)$ read as: 
	\begin{align}
	&\overline{\beta^{2}}=\frac{4n_{\beta}+2p+1}{4q},\nonumber\\
	&\overline{\beta^{-2}}=\frac{4q}{2p-1},\nonumber\\
	&\overline{\beta^{4}}=\frac{24n_{\beta}^2+24pn_{\beta}+4p^2+8p+12n_{\beta}+3}{16q^2},\nonumber\\
	&\overline{\beta^{-4}}=\frac{16q^2\left(4n_{\beta}+2p+1\right)}{\left(2p-3\right)\left(4p^2-1\right)}.
	\label{E37}
	\end{align}
	Here, it should be noted that the analytical calculations of the mean values $\overline{\beta^{t}}$ for large $|t|$($|t|>9$) become rather complicated.
	
		

\section{Reference}
	\begin{thebibliography}{99}
		\bibitem{b1} F. Iachello, Phys. Rev. Lett. 85 (2000) 3580.
		\bibitem{b2} F. Iachello, Phys. Rev. Lett. 87 (2001) 052502.
		\bibitem{b3} F. Iachello, Phys. Rev. Lett. 91 (2003) 132502 .
		\bibitem{b4} P. Cejnar, J. Jolie  and R. F.Casten, Rev. Mod. Phys. 82 (2010) 2155 .
		\bibitem{b5} D. Bonatsos, D. Lenis, D. Petrellis et al, Phys. Lett.B 632 (2006) 238 .
		\bibitem{b6}  D. Bonatsos, D. Lenis, N. Minkov, D. Petrellis et al, Phys. Rev. C 70 (2004) 024305. 
		\bibitem{b7} L. Fortunato, Phys. Rev.C 70 (2004) 011302(R).
		\bibitem{b8} N. Pietralla and O.M. Gorbachenko, Phys. Rev.C 70 (2004) 011304(R).
		\bibitem{b9} F. Iachello and A. Arima, The Interacting Boson Model (Cambridge University Press, Cambridge, England, 1987).
		\bibitem{b10} A. Bohr, Mat. Fys. Medd. Dan. Vid. Selsk. 26 (1952) 14.
		\bibitem{b11} A. Bohr, B. Mottelson, Mat. Fys. Medd. Dan. Vid. Selsk. 27 (1953) 16.
		\bibitem{b12} R. F. Casten and E. A. McCutchan, J. Phys. G 34 (2007) R285 .
		\bibitem{b13} D. Bonatsos, D. Lenis, D. Petrellis and P. A. Terziev, Phys. Lett. B 588 (2004) 172.
		\bibitem{b14} R. F. Casten, Nat. Phys. 2 (2006) 811.
		\bibitem{b15} M. Chabab, A. El Batoul, A. Lahbas and  M. Oulne, Phys. Lett. B 758 (2016) 212.
		\bibitem{b16} M. Alimohammadi and H. Hassanabadi, Nucl. Phys. A 957 (2017) 439.
		\bibitem{b17} M. Alimohammadi, H. Hassanabadi and H. Sobhani, Mod. Phys. Lett. A 31 (2016) 1650193.
		\bibitem{b18} M. Chabab, A. El Batoul, A. Lahbas and M. Oulne, J. Phys. G: Nucl. Part. Phys. 43 (2016) 125107.
		\bibitem{b19} M. Chabab, A. El Batoul, A. Lahbas and  M. Oulne, Nucl. Phys. A 953 (2016)158.
		\bibitem{b20} R. Budaca and A. I. Budaca,  Phys. Rev. C 94 (2016) 054306.
		\bibitem{b21} M. Chabab, A. Lahbas and  M. Oulne, Eur. Phys. J. A 51 (2015) 131.
		\bibitem{b22} M. Chabab, A. Lahbas and  M. Oulne, Phys. Rev. C 91 (2015) 064307.
		\bibitem{b23} R. Budaca, Eur. Phys. J. A 50 (2014) 87.
		\bibitem{b24} R. Budaca, Phys. Lett. B 739 (2014) 56.
		\bibitem{b25} R. Budaca, P. Buganu, M. Chabab et al, Ann. Phys. (NY) 375 (2016) 65.
		\bibitem{b26} D.J. Gross and P.F. Mende, Nucl. Phys. B 303 (1988) 407.
		\bibitem{b27} D. Amati, M. Ciafaloni and G. Veneziano, Nucl. Phys. B 216 (1989) 41.
		\bibitem{b28} K. Konishi, G. Paffuti and P. Provero, Nucl. Phys. B 234 (1990) 276.
		\bibitem{b29} C. Mead, Phys. Rev. 135 (1964) 849.
		\bibitem{b30} L.J. Garay, Int. J. Mod. Phys. A 10 (1995) 145 .
		\bibitem{b31} C. Rovelli and L. Smolin, Nucl. Phys. B 442 (1995) 593.
		\bibitem{b32} A. Sitenko and V. Tartakovskii, Lectures on the Theory of the Nucleus(Atomizdat, Moscow, 1972).
		\bibitem{b33} P. Buganu and R. Budaca, Phys. Rev. C 91 (2015) 014306 .
		\bibitem{b34} F. Brau, J. Phys. A : Math. Gen. 32 (1999) 7691.
		\bibitem{b35} H. Ciftci , R. L. Hall  and N. Saad, J. Phys. A: Math. Gen. 36 (2003)11807 .
		\bibitem{b36} M. Chabab, A. El Batoul and M. Oulne, J. Math. Phys. 56 (2015) 062111.
                \bibitem{b36_ad} D. Bonatsos, P.E. Georgoudis, N. Minkov et al, Phys. Rev. C 88 (2013) 034316.
		\bibitem{b37} I.S. Gradshteyn and I.M.Ryzhik, Table of Integral, Series, and Products (New York: Academic 1980).
		\bibitem{b38} D. Bonatsos, P.E. Georgoudis, D. Lenis et al, Phys. Rev. C 83 (2011) 044321.
		\bibitem{b39} M. Chabab, A. El Batoul, M. Hamzavi et al, Eur. Phys. J. A 53 (2017) 157.
		\bibitem{b40} I. Yigitoglu and M. Gokbulutb, Eur. Phys. J. Plus (2017) 132 .
	        \bibitem{b41} J. Bonnet, A. Krugmann, J. Beller, N. Pietralla, R. V. Jolos, Phys. Rev. C 79 (2009) 034307  .
		\bibitem{b42} Nuclear Data Sheets (\href{http://nndc.bnl.gov/}{http://nndc.bnl.gov/}).
	\end{thebibliography}
\end{document}